\documentclass[twocolumn]{aa}

\usepackage{txfonts}
\usepackage[english]{babel}
\usepackage{amsmath,amssymb}
\usepackage{graphicx, subfig}
\usepackage[normalem]{ulem}
\usepackage{verbatim}
\usepackage{multirow}
\usepackage{mathtools}
\usepackage{epstopdf}
\usepackage{url}
\usepackage{xcolor,color}
\usepackage{orcidlink}
\usepackage{natbib,twoopt}
\usepackage[hyphenbreaks]{breakurl}
\usepackage{booktabs}
\usepackage{placeins}
\usepackage{mhchem}
\usepackage{siunitx}
\usepackage{cleveref}
\usepackage{xspace}
\usepackage{float}
\usepackage{csquotes}
\usepackage{xcolor}

\bibpunct{(}{)}{;}{a}{}{,}             
\definecolor{cobalt}{rgb}{0.06, 0.2, 0.65}
\hypersetup{
  colorlinks=true,
  citecolor=cobalt,
  linkcolor=[rgb]{0.8, 0.2, 1.0},
  urlcolor=cobalt,
  filecolor=magenta, 
}
\makeatletter
  \newcommandtwoopt{\citeads}[3][][]{\href{http://adsabs.harvard.edu/abs/#3}%
    {\def\hyper@linkstart##1##2{}%
     \let\hyper@linkend\@empty\citealp[#1][#2]{#3}}}
  \newcommandtwoopt{\citepads}[3][][]{\href{http://adsabs.harvard.edu/abs/#3}%
    {\def\hyper@linkstart##1##2{}%
     \let\hyper@linkend\@empty\citep[#1][#2]{#3}}}
  \newcommandtwoopt{\citetads}[3][][]{\href{http://adsabs.harvard.edu/abs/#3}%
    {\def\hyper@linkstart##1##2{}%
     \let\hyper@linkend\@empty\citet[#1][#2]{#3}}}
  \newcommandtwoopt{\citeyearads}[3][][]%
    {\href{http://adsabs.harvard.edu/abs/#3}
    {\def\hyper@linkstart##1##2{}%
     \let\hyper@linkend\@empty\citeyear[#1][#2]{#3}}}
\makeatother

\def\apjl{ApJL}
\def\apjs{ApJS}
\def\aap{A\&A}

\linenumbers
\renewcommand\makeLineNumber{}

\graphicspath{{plots/}}
\newcommand{\Msun}{\rm M_\odot}

\definecolor{burgundy}{RGB}{144, 0, 32}

\title{The birth and fate of stellar clumps in super-early galaxies}
\titlerunning{The birth and fate of stellar clumps in super-early galaxies}

\author{
B. Das\inst{1}\fnmsep\thanks{\href{mailto:barnali.das@sns.it}{barnali.das@sns.it}}
\and A. Ferrara\inst{1}
\and A. Pallottini\inst{2}
\and E. Ntormousi\inst{1}
\and M. Kohandel\inst{3, 1}
}           
\authorrunning{Das et al.}

\institute{
Scuola Normale Superiore, Piazza dei Cavalieri 7, 56126 Pisa, Italy
\and 
Dipartimento di Fisica 'Enrico Fermi', Universit\`{a} di Pisa, Largo Bruno Pontecorvo 3, 56127 Pisa, Italy
\and
INAF/OAS Bologna, Via Piero Gobetti 101 / Via Gobetti 93/3, 40129 Bologna, Italy
}

\abstract{
JWST has revealed extremely compact stellar clumps in galaxies at $z>6$, but their formation mechanism and subsequent evolution remain uncertain. We investigate whether such systems can form in-situ through the fragmentation of an early galactic disk and how stellar feedback regulates their properties.
Using RAMSES-RT, we perform a suite of radiation-hydrodynamical simulations of an isolated galaxy ("Ninfea\_blu"), with $\rm M_{vir}=1.5\times10^{10}\,\Msun$, evolved for $100\,{\rm Myr}$ from $z=16$ to $z\simeq12$ with a maximum spatial resolution of $3.6\,{\rm pc}$. The suite explores a variety of feedback processes: thermal, kinetic, photoionization, photoheating, and direct radiation pressure.
In all cases, the gas rapidly forms a rotationally supported disk, reaches a peak star formation rate of $10-15\,\Msun\,{\rm yr}^{-1}$, and fragments into dense stellar clumps. Toomre-unstable regions ($Q_{\rm gas}<1$) appear before the onset of star formation, and the first stellar structures form preferentially within these regions, supporting a Toomre-like gravitational fragmentation pathway.
The clumps have stellar masses of $10^6-2\times10^8\,\Msun$, effective radii of $7-50\,{\rm pc}$, and surface densities of $(0.2-3)\times10^4\, \Msun\,{\rm pc}^{-2}$, overlapping much of the parameter space occupied by observed $z>6$ clumps.
Thermal feedback is strongly affected by radiative losses and produces results close to the no-feedback run. Kinetic feedback reduces the cumulative stellar mass by $30-40\%$, lowers the gas mass retained within clumps by $\approx 1$ dex, and suppresses their high-mass tail without preventing their initial formation. 
Clumps subsequently migrate, interact, undergo tidal stripping, and disperse. In the fiducial run with all the feedback implementations, the clump mass function produces a power law slope of $-1.88$, which aligns closely with observations. Moreover, their contributions to the stellar mass and intrinsic UV luminosity decline from $0.41$ and $0.40$ at $20\,{\rm Myr}$ to $0.13$ and $0.10$ at $100\,{\rm Myr}$, respectively.
We conclude that fragmentation of compact, gas-rich disks provides a viable origin for many of the dense stellar systems observed in the early Universe, while feedback and internal dynamics primarily regulate their growth and fate.
}

\begin{document} 

\maketitle
\section{Introduction}

One of the main objectives of modern astrophysics is to understand how galaxies in the early Universe assembled their stellar mass.
With the advent of powerful telescopes like the James Webb Space Telescope (JWST), a surprisingly over-abundant population of luminous, massive, and blue galaxies at $z>10$, also called ``Blue-monsters" \citep{Ziparo_Ferrara_Sommovigo_Kohandel_2023} was unveiled. Moreover, with the help of gravitational lensing, JWST pushed our capacity to resolve the internal structure of galaxies from $z \lesssim 3$ to $z \sim 12$, revealing individual star forming structures. These massive stellar systems, or stellar clumps, appear to be a ubiquitous feature of star forming galaxies across cosmic time.
At lower redshifts, observations showcase prominent clumpy systems such as the Sparkler at $z=1.4$ \citep{2025A&A...699A.240T}, the Sunburst Arc at $z=2.4$ \citep{2017A&A...608L...4R}, and the Relic at $z=2.5$ \citep{2026ApJ..1001..107W}. However, the discovery of high redshift clumpy galaxies such as the Sunrise arc at $z=5.9$ \citep{2023ApJ...945...53V}, the Cosmic Spear at $z=6.2$ \citep{Abdurrouf_Coe_Resseguier_Murphy_Xu_Adamo_Roy_Henry_Kokorev_Brammer_et_al._2025}, the Firefly galaxy at $z=8.3$ \citep{Mowla_Iyer_Asada_Desprez_Tan_Martis_Sarrouh_Strait_Abraham_Bradač_et_al_2024}, the Cosmic gems at $z=9.6$ \citep{Adamo_Bradley_Vanzella_Claeyssens_Welch_Diego_Mahler_Oguri_Sharon_Abdurrouf_et_al_2024}, the BulletArc-z11 at $z=11.1$ \citep{Bradac_Judez_Willott_Rihtarsic_Martis_Harshan_Felicioni_Asada_Desprez_Clowe_et_al_2025} and the Misty moon at $z= 11-12$ \citep{Nakane_Kokorev_Fujimoto_Ouchi_McLeod_Golubchik_Oguri_Zitrin_Bondestam_Donnan_et_al_2025} demonstrate that star formation in dense and compact clumps extend into the earliest epochs of galaxy formation.

The Blue Monsters at $z>10$ have ($M_{UV} \lesssim -20$) and have stellar masses as massive as $\approx 10^9 \Msun$. Several studies like \cite{Naidu_2022, Xu_2024, Ono_2025} suggest that these galaxies have disk-like morphologies, current resolution limits make it difficult to determine whether such objects have clumpy structures. To address this, attention is often turned to lower redshifts where the combination of ALMA and JWST results provide clear evidence that rotationally supported disks are prevalent and can be susceptible to disk fragmentation \citep{2018Natur.553..178S,2020Natur.584..201R,2021MNRAS.507.3952R, Rizzo_23, 2022ApJ...938L...2F,  2023MNRAS.521.1045R, 2025MNRAS.543.3249D}. A striking example is the Cosmic Grapes at $z=6.1$, which is a set of 15 star forming clumps in a rotating disk \citep{Fujimoto_Ouchi_Kohno_Valentino_Gimenez-Arteaga_Brammer_Furtak_Kohandel_Oguri_Pallottini_et_al_2025}. These systems also show that about 87-97\% of star formation in a galaxy occurs in-situ, that is, the stellar mass grows by accretion rather than mergers \citep{Pushkas_et_al_2025}. Moreover, observations of $z>9$ galaxies with an effective radius in the range 150 to 650 pc sustain a high star formation rate of $\sim 10 \, \Msun \, \rm yr^{-1}$ \citep{Ferrara_Rodighiero_Carniani_Zhang_Kohandel_Das_2026}.

The $z>6$ clumps in galaxies where star formation is most likely the primary power source, have stellar masses of order $10^6 \, \Msun$ and stellar surface densities as high as $10^4-10^6 \, \Msun \, \rm pc^{-2}$ which are an order of magnitude higher than the local globular clusters (GC) and several orders of magnitude higher than the young star clusters (YSC; \citealt{Brown_Gnedin_2021}) found in local universe \citep{Claeyssens_Adamo_Kokorev_Furtak_Richard_Beauchesne_Dessauges-Zavadsky_Atek_Chisholm_Endsley_et_al_2026}. Due to the expected high gas density in the clumps, stellar feedback becomes inefficient, potentially boosting the local star formation efficiency to values as high as 0.6-0.9 \citep{Fujimoto_Ouchi_Kohno_Valentino_Gimenez-Arteaga_Brammer_Furtak_Kohandel_Oguri_Pallottini_et_al_2025}. Moreover, they are the major producer of UV light in the galaxy. In cases like Misty Moon, the stellar mass budget is over 80\% of the total mass of the galaxy, while the BulletArc-z11 contains approximately 50\% of the total stellar mass even though the clump sizes are $\lesssim$ 10 pc. 

How these extremely compact clumps form remains an open question, with proposed mechanisms broadly divided into in-situ and ex-situ pathways.
On one hand, in the in-situ scenario, clumps form within the host galaxy due to internal dynamics. At high redshifts, disk instabilities can occur as rapid cold gas accretion and extreme gas density force unstable disks to break apart into clumps due to Toomre instability, causing Violent Disk Instability (VDI) \citep{2009ApJ...703..785D, Bournaud_Elmegreen_Elmegreen_2007, Agertz_Teyssier_Moore_2009,2004A&A...413..547I}. In-situ clumps can also form due to turbulence-driven fragmentation where turbulence in dense gas creates local overdensities that collapse into clumps \citep{2025MNRAS.538L...9M, 2025A&A...698A.110G}, filament fragmentation \citep{2023MNRAS.522.2495G, 2025OJAp....8E.146G}, where the infalling cold and dense gas filaments form stellar clumps before reaching the disk which later orbit inward or due to radiation-regulated collapse \citep{2024ApJ...970...14S} where strong FUV radiation prevents star formation by suppressing cooling, thereby forcing gas to collapse into stars suddenly. The clumps inherit the disk angular momentum, are almost coeval, and endure secular processes within the galaxy. 
On the other hand, ex-situ pathways describe clumps that form due to external factors like mergers, satellite capture, or tidal shocks \citep{2019A&A...632A..98C,van_Donkelaar_Mayer_Capelo_Sijacki_Adamo_2026}. These ex-situ clumps are expected to show diverse ages, metallicities, and orbital configurations \citep{Zanella_LeFloch_Harrison_Daddi_Bernhard_Gobat_Strazzullo_Valentino_Cibinel_SanchezAlmeida_et_al._2019, Nakazato_Ceverino_Yoshida_2024, Gutcke_2024}. In the high redshift environment, where both mergers and rotationally supported disks are prevalent \citep{2018Natur.553..178S,2020Natur.584..201R,2021MNRAS.507.3952R, 2022ApJ...938L...2F,Rizzo_23, 2023ApJ...955...94F, 2023MNRAS.521.1045R, 2025MNRAS.543.3249D}, one can expect both scenarios to be at play. 

However, with the help of numerical simulations, it is possible to predict the most important pathway that gives rise to the compact clumps that are observed. Previously, various simulation studies have been undertaken to understand how these compact star clusters emerge and evolve. Cosmological zoom-in simulations at high redshifts do show clumpy structures in galaxies, contended to be formed via in-situ \citep{Katz_Galligan_Kimm_Rosdahl_Haehnelt_Blaizot_Devriendt_Slyz_Laporte_Ellis_2019, Pallottini_Ferrara_Gallerani_Behrens_Kohandel_Carniani_Vallini_Salvadori_Gelli_Sommovigo_etal_2022} and ex-situ \citep{Nakazato_Ceverino_Yoshida_2024} pathways. There are also a few sub parsec resolution simulations, out of which \cite{Garcia_Ricotti_Sugimura_Park_2023} shows that the clumps form through filament fragmentation. Another ex-situ pathway is illustrated by \cite{Gutcke_2024} where star clusters form in independent low mass DM halos and transform into present day baryon dominated star clusters (resembling GCs) as the DM is tidally stripped after they are being accreted into larger host dwarf galaxies. Other studies \citep{Ma_Grudic_Quataert_Hopkins_Faucher-Giguere_Boylan-Kolchin_Wetzel_Kim_Murray_Keres_2020,Mayer_VanDonkelaar_Messa_Capelo_Adamo_2025,2025ApJ...990..135W, Calura_Pascale_Agertz_Andersson_Lacchin_Lupi_Meneghetti_Nipoti_Ragagnin_Rosdahl_et_al._2025, Pascale_Calura_Vesperini_Rosdahl_Nipoti_Giunchi_Lacchin_Lupi_Messa_Meneghetti_et_al._2025} show that the clump formation is due to in-situ ways like compression by feedback-driven winds and disk fragmentation. Complementarily, there are also isolated galaxy studies that look into clump properties in a disk galaxy \citep{2018MNRAS.475.4617K, Hirai_Fujii_Saitoh_2021, 2023ApJ...950..132H, Deng_Li_Liu_Kannan_Smith_Bryan_2024}.

While these simulations have provided valuable insight into clump formation, they share important limitations that hinder the proper understanding of observed high-z clumps. Cosmological simulations proposing an in-situ mechanism of clump formation, often lack the resolution to probe parsec scale clumps \citep{Katz_Galligan_Kimm_Rosdahl_Haehnelt_Blaizot_Devriendt_Slyz_Laporte_Ellis_2019, Pallottini_Ferrara_Gallerani_Behrens_Kohandel_Carniani_Vallini_Salvadori_Gelli_Sommovigo_etal_2022}, produce morphologies without well defined disks \citep{Garcia_Ricotti_Sugimura_Park_2023}, cannot fully separate the influence of black holes and merger histories \citep{Mayer_VanDonkelaar_Messa_Capelo_Adamo_2025}, adopt a setup without stellar feedback \citep{2025ApJ...990..135W} or rely on extremely high star formation efficiencies to reproduce observed clump densities \citep{Calura_Pascale_Agertz_Andersson_Lacchin_Lupi_Meneghetti_Nipoti_Ragagnin_Rosdahl_et_al._2025, Pascale_Calura_Vesperini_Rosdahl_Nipoti_Giunchi_Lacchin_Lupi_Messa_Meneghetti_et_al._2025}. 
Overall, in cosmological simulations, one cannot entirely disentangle the environmental effect, but on the other hand, the isolated simulation studies assume the galaxy to have a pre-existing stellar disk from the start, thereby disregarding the effect of feedback from the initial stars, which can play an important role in clump properties. This motivates the use of isolated cases of galaxy formation which naturally form stars that inject feedback from the beginning, allowing a clean investigation of in-situ clump formation. 

Clumps in the above simulations are affected by stellar feedback and thereby modify the star formation history of the entire galaxy. Feedback due to supernova explosions and radiation pressure can disperse gas and alter clump densities and their survivability \citep{Menon_Federrath_Krumholz_2023, 2025MNRAS.537.1646N}. As different feedback mechanisms act on different spatial and temporal scales with different efficiencies, the clump properties may vary across simulation implementations. This calls for exploring how various feedback mechanisms affect clump properties in an isolated high redshift galaxy.

In this paper, we investigate whether a galaxy can naturally form a clumpy disk at $z=16$. We then follow its evolution down to $z=12.5$ under different feedback implementations and analyze how these feedback channels affect the overall galaxy and the properties of the clumps. This redshift window corresponds to the epoch where the observed very high$-z$ galaxies are expected to be unstable and form clumps \citep{2026arXiv260917667F} but remain unresolved by JWST, allowing us to predict their clumpy structures.

In Sec.~\ref{sec:Setup}, we describe the numerical setup, initial
conditions, and stellar-feedback prescriptions. In
Sec.~\ref{sec:evol}, we present the global evolution of the
simulated galaxy, including its morphology, star-formation
history, outflows, internal kinematics, and gravitational
stability. In Sec.~\ref{sec:clumps}, we characterize the
identified stellar clumps, investigate their dynamics and UV
contribution, and compare their properties with observed compact
stellar systems. In Sec.~\ref{sec:discussion}, we discuss the
physical interpretation, observational implications, and
limitations of our results. Finally, Sec.~\ref{sec:summary}
summarizes our main conclusions and outlines directions for
future work.

\section{Numerical Setup}\label{sec:Setup}

To simulate isolated galaxies, we use the publicly available Adaptive Mesh Refinement (AMR) code RAMSES-RT \citep{Rosdahl_Blaizot_Aubert_Stranex_Teyssier_2013, Rosdahl_Teyssier_2015, Rosdahl_Schaye_Teyssier_Agertz_2015}. RAMSES-RT treats the DM and stars as collisionless particles, calculating their self-gravity using a multigrid particle-mesh solver \citep{2011JCoPh.230.4756G}. The gas is treated as a fluid and is evolved using a second-order Godunov hydrodynamics scheme. 

We follow the emergence of clumps in a controlled environment for a single galaxy, called "Ninfea\_blu", in which stars form from gas density peaks, and we adopt five different feedback prescriptions.
Although the SN thermal feedback is a simple and natural choice, it is highly susceptible to numerical overcooling depending on the resolution \citep{2013MNRAS.429.1922C}. For this reason, we also explore the SN kinetic feedback with the expectation that the physical SN feedback should lie between the purely thermal and purely kinetic limits. We also include feedback due to radiation from stars in the form of photoionization and radiation pressure. The simulation runs and their feedback prescription are summarized in Table \ref{tab:feedback_models}. 

\begin{table}
\centering 
\caption{Feedback physics included in each simulation run. All runs share the same star-formation model. Note that in the \texttt{No\_fb} (no feedback) run, SNs eject metals but not energy.
\label{tab:feedback_models} 
} 
\begin{tabular*}{\columnwidth}{@{\extracolsep{\fill}}l ccc} 
\toprule & \multicolumn{2}{c}{Supernova} & {Photoionization +} \\ 
\cmidrule(lr){2-3} 
Run & Thermal & Kinetic & Radiation pressure\\
\midrule \texttt{No\_fb} & $\times$ & $\times$ & $\times$ \\
\texttt{Th\_fb} & $\checkmark$ & $\times$ & $\times$ \\
\texttt{ThR\_fb} & $\checkmark$ & $\times$ & $\checkmark$\\
\texttt{Kn\_fb} & $\times$ & $\checkmark$ & $\times$ \\
\texttt{KnR\_fb} & $\times$ & $\checkmark$ & $\checkmark$ \\
\bottomrule 
\end{tabular*} 
\end{table}

Below we describe the numerical methods, initial conditions, resolution strategies, and feedback models used in our simulations. 

\subsection{Initial setup}

Well established theories \citep[see][for reviews]{2010gfe..book.....M, Dayal_Ferrara_2019} suggest that isolated galaxies can form via hot- or cold-mode accretion depending on their masses and redshift. In halos with virial mass $>10^{11.5} \, \Msun$, gas falls into the dark matter halo and is shock heated to the virial temperature of the halo and then slowly cools to form a disk, following a hot-mode accretion. On the other hand, smaller halos ($\lesssim 10^{11.5} \, \Msun$) follow the cold-mode accretion scenario where gas is accreted at temperatures much lower than the virial temperature. As we aim to evolve Ninfea\_blu from $z=16$ to $z=12.5$, we adopt a DM halo with a virial mass of $M_h=1.5\times 10^{10}\,\Msun$ (corresponding to a virial velocity $v_{\rm vir}\simeq 121\ \rm km\ s^{-1}$ and virial radius $r_{\rm vir}\simeq 4.4\ \rm kpc$ at $z=16$), a value consistent with the inferred halo mass of the observed "Blue Monsters" \citep{Ferrara_Rodighiero_Carniani_Zhang_Kohandel_Das_2026}. 
To mimic the cold-mode accretion, we also consider a gaseous halo at $\sim \SI{1e4}{K}$, which is the atomic cooling threshold. We use the open source code Disk Initial Conditions Environment (DICE) \citep{2016ascl.soft07002P} to setup the DM and gas components\footnote{We assume a flat Universe with the following cosmological parameters: $\Omega_m = 0.30$, $\Omega_{\Lambda} = 1- \Omega_{\rm M}$, and $\Omega_{b} = 0.0486$,  $h=0.71$, where $\Omega_{m}$, $\Omega_{\Lambda}$, and $\Omega_{b}$ are the total matter, vacuum, and baryon densities, in units of the critical density; $h$ is the Hubble constant in units of $100\,\rm km\ s^{-1} Mpc^{-1}$.}. The baryon fraction is 5\% of the virial mass. The mass resolution for DM and gas is $7.31 \times 10^3 \, \Msun$ and $3.58 \times 10^2 \, \Msun$, respectively. 

The dark matter halo density profile follows an NFW fit \citep{1997ApJ...490..493N} with concentration $c=5$, while the gas distribution is less concentrated to account for pressure effects, $c=2$. The spin parameter is taken to be $\lambda = 0.016$, which lies within the expected lognormal distribution of $\langle \lambda \rangle \simeq 0.03$ with $\sigma_{\ln \lambda} \simeq 0.5$ \citep{2007MNRAS.378...55M}. We adopt this specific value as it matches the spin measured for the Amaryllis, a cosmological zoom-in galaxy \citep{Kohandel_Pallottini_Ferrara_2025} which reproduces some of the properties of high-z observed galaxies, enabling future one-to-one comparison, although such a comparison is beyond the scope of this paper. For this choice of spin, a baryons to DM specific angular momentum ratio of $\sim 4$ has been chosen to obtain a disk radius comparable to that of Blue Monsters. We assume an initially primordial ($Z=0$) gas composition.

\begin{table*}
\centering
\caption{Observed properties of spectroscopically confirmed galaxies at
$z \geq 12$.}
\label{tab:observed_galaxies}
\begingroup
\small
\renewcommand{\arraystretch}{1.15} 
\setlength{\tabcolsep}{4pt}        
\begin{tabular}{@{}lcccccl@{}}
\toprule
Galaxy ID &
$z$ &
SFR [$\Msun\,{\rm yr}^{-1}$] &
$\log(M_\star/\Msun)$ &
$R_{\rm eff}$ [pc] &
Age [Myr] &
Reference \\
\midrule

GHZ2 &
$12.34$ &
$5.20^{+1.10}_{-0.60}$ &
$9.05^{+0.10}_{-0.25}$ &
$105^{+9}_{-9}$ &
$28^{+10}_{-14}\,^\star$ &
\cite{2024ApJ...972..143C} \\

UNCOVER-z12 &
$12.39$ &
$2.15^{+0.81}_{-0.46}$ &
$8.35^{+0.14}_{-0.18}$ &
$426^{+40}_{-42}$ &
$61.66^{+31.67}_{-18.01}\,^\dagger$ &
\cite{2023ApJ...957L..34W} \\

GS-z12-0 &
$12.63$ &
$1.80^{+0.54}_{-0.43}$ &
$7.64^{+0.66}_{-0.39}$ &
$144^{+15}_{-15}$ &
$22.91^{+105.92}_{-17.02}\,^\dagger$ &
\cite{2023NatAs...7..622C} \\

GS-z13-1-LA &
$13.01$ &
$0.16^{+0.48}_{-0.15}$ &
$7.74^{+0.52}_{-0.18}$ &
$14^{+14}_{-7}$ &
$22^{+9}_{-6}\,^\star$ &
\cite{2025Natur.639..897W} \\

UNCOVER-z13 &
$13.08$ &
$1.28^{+0.27}_{-0.18}$ &
$8.13^{+0.11}_{-0.15}$ &
$309^{+110}_{-74}$ &
$66.1^{+25.1}_{-21.4}\,^\dagger$ &
\cite{2023ApJ...957L..34W} \\

GS-z13-0 &
$13.20$ &
$1.36^{+0.31}_{-0.23}$ &
$7.95^{+0.19}_{-0.29}$ &
$<52$ &
$69.2^{+48.3}_{-39.0}\,^\dagger$ &
\cite{2023NatAs...7..622C} \\

PAN-z14-1 &
$13.53$ &
$4.80^{+13.00}_{-4.80}$ &
$8.23^{+1.14}_{-0.21}$ &
$233^{+10}_{-10}$ &
$5^{+61}_{-4}\,^\star$ &
\cite{2026ApJ..1002..134D} \\

GS-z14-1 &
$13.90$ &
$2.00^{+0.70}_{-0.40}$ &
$8.00^{+0.40}_{-0.30}$ &
$<160$ &
$<20\,^\star$ &
\cite{2024Natur.633..318C} \\

GS-z14-0 &
$14.18$ &
$19.00^{+6.00}_{-6.00}$ &
$8.84^{+0.09}_{-0.10}$ &
$260^{+2}_{-2}$ &
$40^{+5}_{-5}\,^\star$ &
\cite{Carniani25} \\

MoM-z14 &
$14.44$ &
$13.00^{+3.70}_{-3.50}$ &
$8.10^{+0.30}_{-0.20}$ &
$74^{+15}_{-12}$ &
$4.0^{+10.0}_{-1.4}\,^\ddagger$ &
\cite{2026OJAp....956033N} \\
\bottomrule
\end{tabular}
\smallskip
\begin{minipage}{\textwidth}
\footnotesize
\hspace{2.4cm} 
$^\star$Mass-weighted stellar age;\quad
$^\dagger$Age of the oldest stars;\quad
$^\ddagger$Half-mass stellar assembly timescale.
\end{minipage}
\endgroup
\end{table*}

\subsection{Adaptive refinement}

The AMR grid refines according to both the local gas density and the Jeans criterion, ensuring the local Jeans length is resolved by at least four cells. The simulation box length is 30 kpc, with the maximum and minimum cell width being 117.2 pc and 3.6 pc, respectively.

\subsection{Gas Thermodynamics}

The cooling mechanism is handled by a non-equilibrium thermochemistry module that evolves the ionization states of hydrogen and helium ions and molecule (HI, HII, H$_2$, HeI, HeII and HeIII) together with the local radiation field (from stars; if applied) \citep{Nickerson_Teyssier_Rosdahl_2018}. The gas cools because of a combination of the following processes: collisional excitation, ionization, recombination and bremsstrahlung, molecular H$_2$ (at low temperatures) and metal line cooling using CLOUDY-based tables \cite{1998PASP..110..761F}. Heating sources mainly include thermal energy generated by SNe and photoionization from stellar radiation. 

\subsection{Star Formation}

Stars form when the gas temperature drops below $T_{\rm crit}=500$ K, and its density in a cell exceeds the threshold of $n_{\rm crit}=1500\ \rm cm^{-3}$, consistent with observations of local star-forming molecular clouds. The temperature criterion ensures that stars form only in very cold gas as expected when H$_2$ cooling becomes efficient in high-$z$ environments. Following the standard Schmidt law \citep{1959ApJ...129..243S, 1998ApJ...498..541K}, the gas is converted into stars at a rate of
\begin{equation}
\dot{\rho_\star} = \epsilon_{\rm ff}\rho / t_{\rm ff}\,,,
\end{equation}
where $\rho$ is the gas density and $\epsilon_{\rm ff} = 0.1$ is the star formation efficiency per free fall time, $t_{\rm ff} = [3\pi/(32G\rho)]^{1/2}$, with $G$ being the gravitational constant. The choice of $\epsilon_{\rm ff}$ is the average value observed for the molecular clouds in the Milky Way (MW) \citep{Murray_2011}. Collisionless particles, representing stellar populations, are formed stochastically from the gas, with the probability of forming one drawn from a Poissonian distribution.

\subsection{Supernova Feedback}

Our setup includes both thermal and kinetic supernova feedback, employed in separate simulation runs. In the former, we use the `thermal dump' model which, after 3 Myr of the stellar particle's birth, injects mass $m_{\rm ej} = \eta_{\rm SN} \, m_*$, and thermal energy $\epsilon_{\rm SN} =  10^{51} \,\eta_{\rm SN} \,  (m_* /10 \, M_{\odot})\, \rm erg$ into its host cell; we take  $\eta_{\rm SN} = 0.1$. Supernovae are also assumed to inject a metal mass $m_Z = 0.1 m_{\rm ej}$. 

For the kinetic SN feedback, we adopt \cite{2008A&A...477...79D}. When a star particle reaches the age of 3 Myr, a fraction of its mass ($\eta_{\rm SN}$ = 0.1) is returned to the gas as SN ejecta, with a metal yield of 0.1. The SN energy budget ($10^{51}$ erg per SN‑equivalent) is injected purely in kinetic form. The feedback uses the debris‑particle scheme, which redistributes mass, momentum, and energy within a spherical blast region of radius 20 pc, corresponding to 5-6 cells at our resolution.

\subsection{Radiative feedback}

In runs with RT, the radiation transport is applied on-the-fly using the BC03 Stellar Population Synthesis (SPS) model \citep{2003MNRAS.344.1000B}. The RAMSES-RT code advects photons using a momentum-based framework with M1 closure for the Eddington tensor. The photons interact with hydrogen and helium (HI, HII, H$_2$, HeI, HeII, and HeIII) via photoionization, heating, and momentum transfer. They also impart momentum onto the dust, which moves as a passive scalar advected with metals. We adopt a reduced speed of light of 0.01c for the photon propagation to optimize the computational load. We make sure that this reduced speed is faster than the dissociation and ionization front propagation. We adopt the on-the-spot approximation, corresponding to case B recombination, which is a standard choice for the ISM. The photons are divided into 5 frequency groups: 8.152-11.20 eV, 11.20-13.6 eV (H$_2$ photo-dissociating), 13.60-24.59 eV (HI ionizing), 24.59-54.42 eV (HeI ionizing), 54.42-$\infty$ (HeII ionizing). Their corresponding dust absorption (Planck) opacity factor, assuming a MW like dust extinction \citep{Weingartner_2001}, is given in Tab. \ref{tab:energy_groups}.

\begin{table}
\centering
\caption{Energy ranges and average opacities for different groups (including scattering)}
\label{tab:energy_groups}
\begin{tabular}{lcc}
\toprule
Group & Energy Range (eV) & Avg. $\kappa$ ($\text{cm}^2/\text{g}$) \\
\midrule
Pre-LW   & 08.15 -- 11.20          & $1.01 \times 10^5$ \\
LW       & 11.20 -- 13.60          & $1.49 \times 10^5$ \\
HI Ion   & 13.60 -- 24.59          & $1.67 \times 10^5$ \\
HeI Ion  & 24.59 -- 54.42          & $8.44 \times 10^4$ \\
HeII Ion & 54.42 -- 100.00         & $5.96 \times 10^4$ \\
\bottomrule
\end{tabular}
\end{table}

The galaxy depletion time computed as $t_{\rm dep} = M_{\rm gas}/\rm SFR$, where $M_{\rm gas}$ is the gas mass in the galaxy, and SFR is the star formation rate, is the time required for the galaxy to consume its gas reservoir to form stars, assuming a constant SFR. For Ninfea\_blu, $M_{\rm gas} = 7.5 \times 10^8 \, \Msun$ and we can consider an average SFR of $6\ \, \Msun\ \rm yr^{-1}$, giving $t_{\rm dep} = 125\ \rm Myr$. As simulations diverge from realistic conditions after the depletion time, we run our simulations in RAMSES-RT up to $\SI{100}{Myr}$.

\begin{figure*}
    \centering
    \includegraphics[width=1\linewidth]{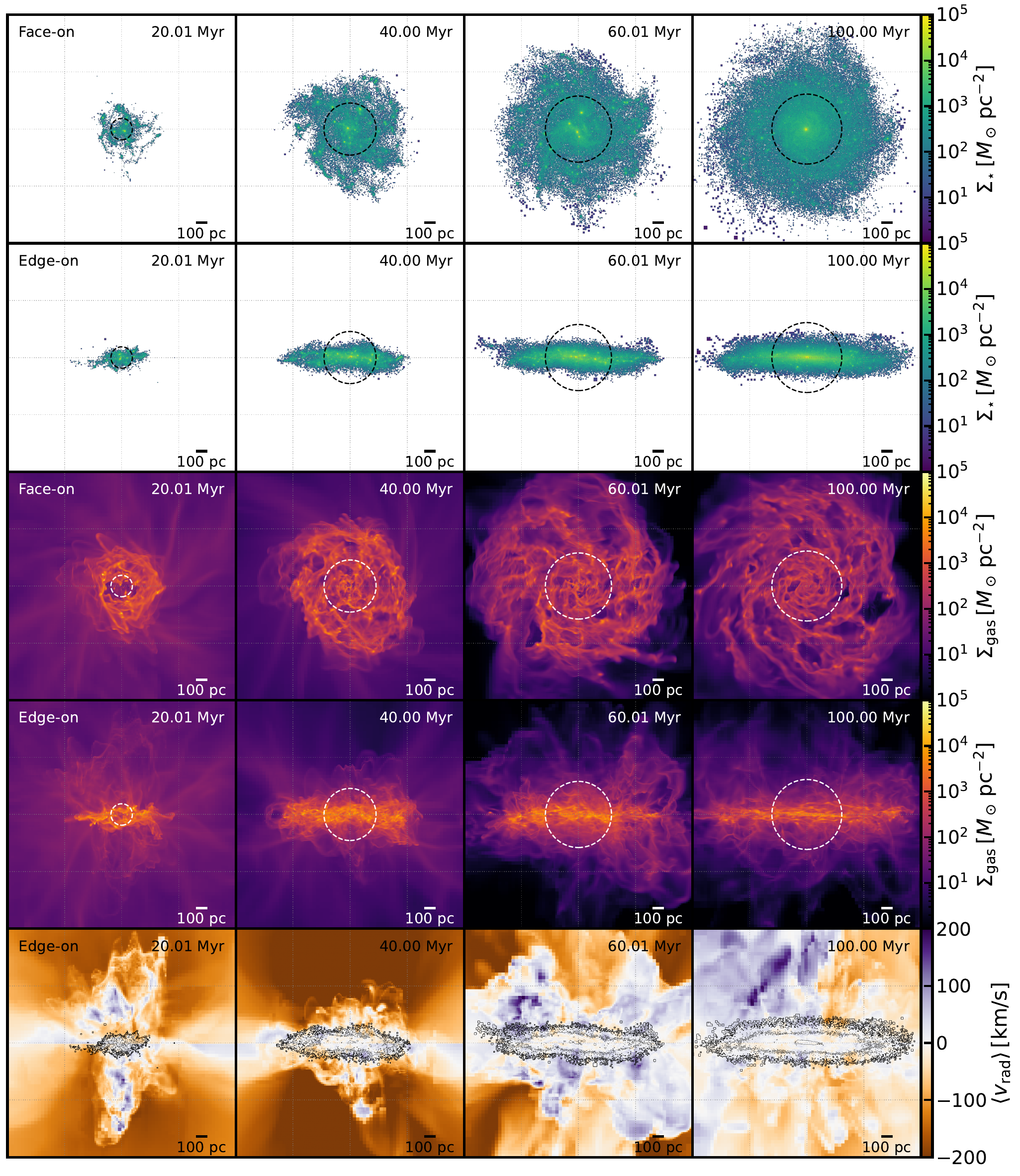}
    \caption{
    Overview of our fiducial run, \texttt{KnR\_fb}. Rows 1-2 show the stellar surface density ($\Sigma_\star$, face-on and edge-on, respectively), rows 3-4 show the gas surface density ($\Sigma_{\rm gas}$, face-on and edge-on, respectively), and row 5 shows the mass-weighted radial gas velocity ($\langle v_{\rm rad}\rangle$, edge-on), where positive and negative values indicate outflowing and inflowing gas, respectively.
    Different columns report the evolution at different times, as indicated in the top right corner of each panel.
    The start of the simulation ($t=0$) corresponds to $z=16$, therefore, the snapshots at 20 Myr, 40 Myr, 60 Myr correspond to redshifts $z = 15.14,\, 14.38,\, 13.71,\, \mathrm{and}\, 12.55$, respectively.
    The spatial scale is shown as an inset.
    The black (white) circles in the $\Sigma_\star$ ($\Sigma_{\rm gas}$) maps mark the stellar half-mass radius of the galaxy as it evolves.
    In row 5, the black contour traces the stellar density distribution. 
    \label{fig:maps}    
    }
\end{figure*}

\section{Global galaxy evolution}
\label{sec:evol}

In the next subsections, we discuss several aspects of Ninfea\_blu's evolution and its overall properties, with particular emphasis on the role of feedback, the presence of clumps, and the formation of a disk. 

\subsection{Morphology and outflows}\label{sec:morph}

Fig.~\ref{fig:maps} provides an overview of the morphological and kinematic evolution of our fiducial run, \texttt{KnR\_fb}, which includes kinetic supernova feedback, photoionization, photoheating, and direct radiation pressure.
The four columns correspond to $t=20$, 40, 60, and $100\,{\rm Myr}$ and therefore sample the rising, peak, and declining phases of the galaxy star formation history. The system undergoes rapid inside-out growth: a compact ($\sim 100$ pc) stellar component is already present after $20\,{\rm Myr}$, but the stellar extent increases continuously thereafter. By $60$--$100\,{\rm Myr}$, the stars form an extended, flattened distribution with a half-mass radius approaching $R_{50}\sim 300\ \rm pc$. The edge-on maps show that this component remains disk-like throughout the evolution, although it becomes progressively thicker and more spatially diffuse. 

The stellar disk is strongly structured rather than smooth. Several compact surface-density peaks are already visible at $20\,{\rm Myr}$ and new overdensities continue to appear as the disk grows. These dense stellar clumps coexist with a lower-density diffuse component produced by star formation outside the main peaks or by tidal stripping from clumps with subsequent redistribution by differential rotation. 
The persistence of multiple compact peaks until $100\,{\rm Myr}$ shows that stellar feedback does not prevent the initial formation of bound or partially bound stellar structures. At the same time, the increasingly diffuse morphology indicates that the clump population is continuously evolving rather than remaining as a fixed collection of isolated systems.

The gas exhibits an even more complex morphology. In the face-on projections, it forms a network of dense, curved filaments and knots embedded in a lower column density medium. These structures trace the material feeding the star-forming disk and provide the dense reservoirs from which the stellar clumps form. The edge-on projections reveal a geometrically thin, high-column-density layer surrounded by a much more extended, low-density extraplanar component. Thus, despite the action of feedback, the cold and dense gas remains concentrated close to the equatorial plane, while lower-density gas is lifted to progressively larger heights. The spatial correspondence between stellar and gaseous overdensities is particularly clear at early times, whereas at later times many stellar clumps contain relatively little dense gas, suggesting that feedback removes gas more efficiently than it disperses the already formed stellar component.

The radial-velocity maps show that this vertical extension is produced by a highly anisotropic, multiphase galactic outflow. Gas is preferentially expelled perpendicular to the disk, where it encounters the smallest column density, while inflowing and outflowing material coexist near the disk and along the walls of the expanding cavities. The flow is most clearly bipolar at $20$--$40\,{\rm Myr}$ and becomes increasingly irregular at later times as successive feedback events interact with previously displaced gas. Outflow velocities locally can reach up to $\simeq 200\,{\rm km\,s^{-1}}$, exceeding the halo virial velocity, $v_{\rm vir}=120.7\,{\rm km\,s^{-1}}$. 

Overall, Fig.~\ref{fig:maps} illustrates the coexistence of three processes governing the evolution of the system: continued assembly of a rotationally supported disk, fragmentation of its dense gas into compact stellar structures, and feedback-driven removal of gas predominantly along the minor axis.

\begin{figure*}
    \centering
    \includegraphics[width=1.0\linewidth]{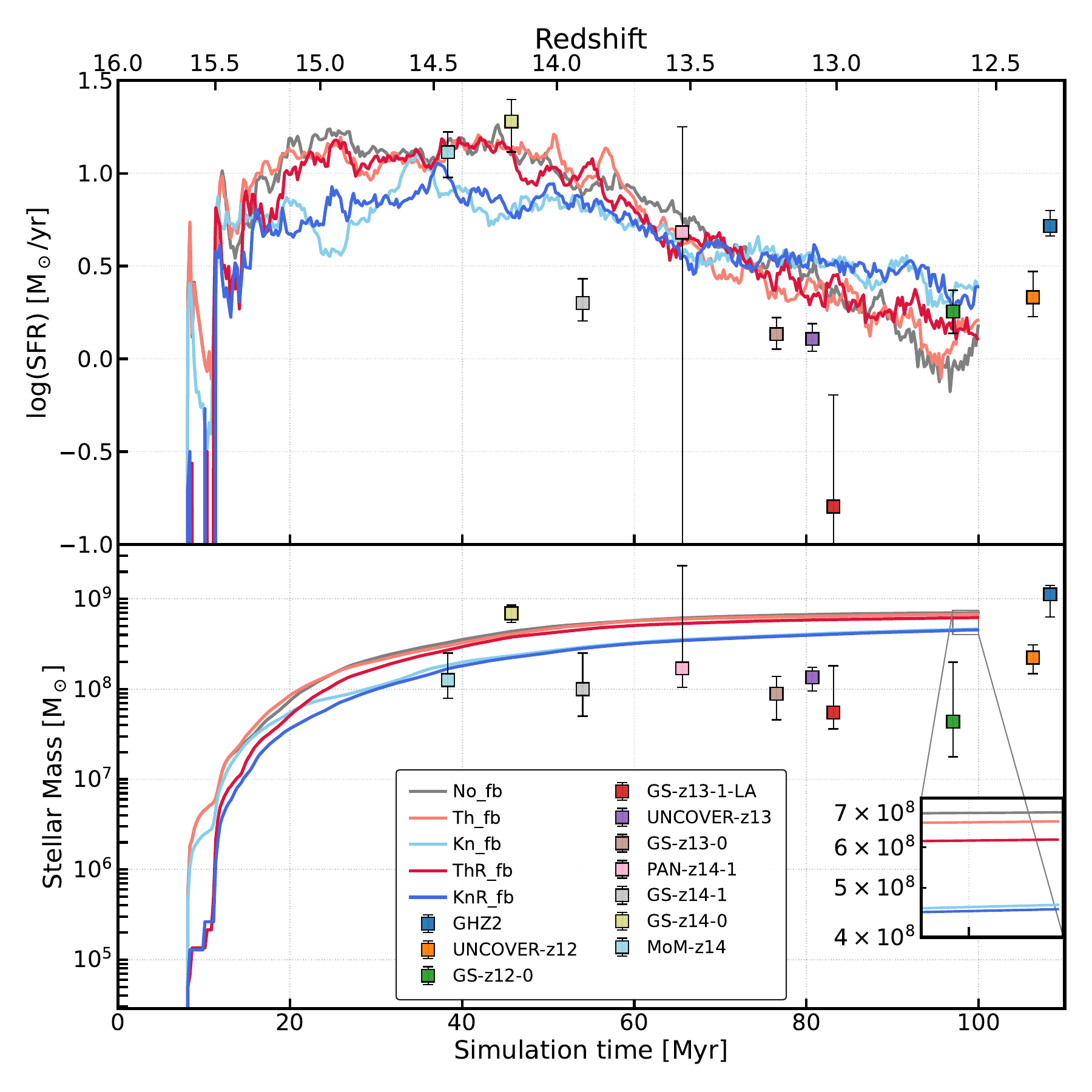}
    \caption{Star formation history (top panel) and stellar mass evolution (bottom) as a function of simulation time and redshift (upper axis) for the five runs, shown as solid lines: \texttt{No\_fb} (gray), \texttt{Th\_fb} (salmon), \texttt{Kn\_fb} (skyblue), \texttt{ThR\_fb} (crimson), \texttt{KnR\_fb} (royalblue). The inset highlights the difference among the runs at the end of the simulation. The square symbols with error bars denote observed high redshift galaxies (see Tab.\ref{tab:observed_galaxies}).}
    \label{fig:sfh}
\end{figure*}

\subsection{Star formation history and feedback}
\label{sec:sfh}

Fig.~\ref{fig:sfh} shows the evolution of the star formation rate and cumulative stellar mass for the five feedback prescriptions explored for Ninfea\_blu. Star formation begins at $t\simeq8\,{\rm Myr}$, once the initially warm gas has cooled and reached the adopted density and temperature thresholds, $n_{\rm crit}$ and $T_{\rm crit}$, respectively. The first episode is highly variable, with a rapid rise followed by a brief minimum at $t\simeq10$--$12\,{\rm Myr}$. This transient behaviour reflects the small amount of gas initially able to reach the star-forming phase and the discrete formation of the first dense structures. Since it occurs during the initial relaxation of the galaxy, we do not attach particular physical significance to its detailed shape.

After this initial phase, the SFR rises rapidly in all runs and reaches approximately $10-15\, \Msun\,{\rm yr}^{-1}$ between $20$ and $50\,{\rm Myr}$. This period corresponds to the assembly and fragmentation of the compact gas disk seen in Fig.~\ref{fig:maps}. The broad similarity of the five histories demonstrates that none of the feedback prescriptions prevents the galaxy from entering an intense star-forming phase. The differences among the runs are nevertheless systematic: models employing kinetic supernova feedback generally have SFRs lower by $\simeq0.2-0.3$ dex than the no-feedback and thermal-feedback runs during the main growth phase.

Beyond $t\simeq50\,{\rm Myr}$, the SFR declines in every model, reaching values of approximately $1-3\,\Msun\,{\rm yr}^{-1}$ by $100\,{\rm Myr}$. This common decline is primarily associated with the progressive exhaustion and redistribution of the finite initial gas reservoir. Because the simulations are isolated, the consumed or expelled gas is not replenished by cosmological accretion. The late evolution should therefore not be interpreted as a general prediction that galaxies of this mass must quench over a timescale of $100\,{\rm Myr}$. In a cosmological environment, continued gas accretion could sustain a high SFR or generate additional star-formation episodes. Quantitative predictions at late times are consequently less robust than the formation and early fragmentation of the disk.

The bottom panel shows the stellar mass build-up process. By $100\,{\rm Myr}$, the no-feedback and thermal-feedback runs have formed approximately $(6-7)\times10^8\,\Msun$ of stars, whereas kinetic feedback limits the final stellar mass to $\simeq(4-5)\times10^8\,\Msun$. Thermal supernova feedback remains close to the no-feedback case because energy deposited in dense gas is rapidly radiated away. By contrast, kinetic injection couples supernova momentum more effectively to the gas and suppresses the total conversion of gas into stars by approximately $30-40\%$. Adding radiative feedback\footnote{We warn that our simulations, while including dust-mediated radiation pressure, do not account for the additional contribution of Ly$\alpha$ radiation pressure which is found to be dominant in (quasi-)primordial environments \citep{Ferrara25_a, Manzoni25, Nebrin25}} produces a significant suppression of star formation during the first 20 Myr, but such effect tends to become weaker and comparable to the thermal feedback at later evolutionary stages. 

The observed $z\geq12$ galaxies listed in Table~\ref{tab:observed_galaxies} span ${\rm SFR}\simeq0.1-20\,\Msun\,{\rm yr}^{-1}$ and $M_\star\simeq5\times10^7-10^9\,\Msun$. The simulated galaxy therefore lies within the observed range, particularly among the brighter and more massive systems. This comparison should be regarded as a consistency check rather than as an evolutionary fit to individual objects: the observations sample galaxies with different halo masses and assembly histories, whereas our curves follow a single, deliberately massive halo from identical initial conditions. Nevertheless, the simulations show that a $1.5\times10^{10}\,\Msun$ halo can simultaneously reproduce the high SFRs and large stellar masses inferred for some of the earliest spectroscopically confirmed galaxies.

\begin{figure}
    \centering
    \includegraphics[width=1\linewidth]{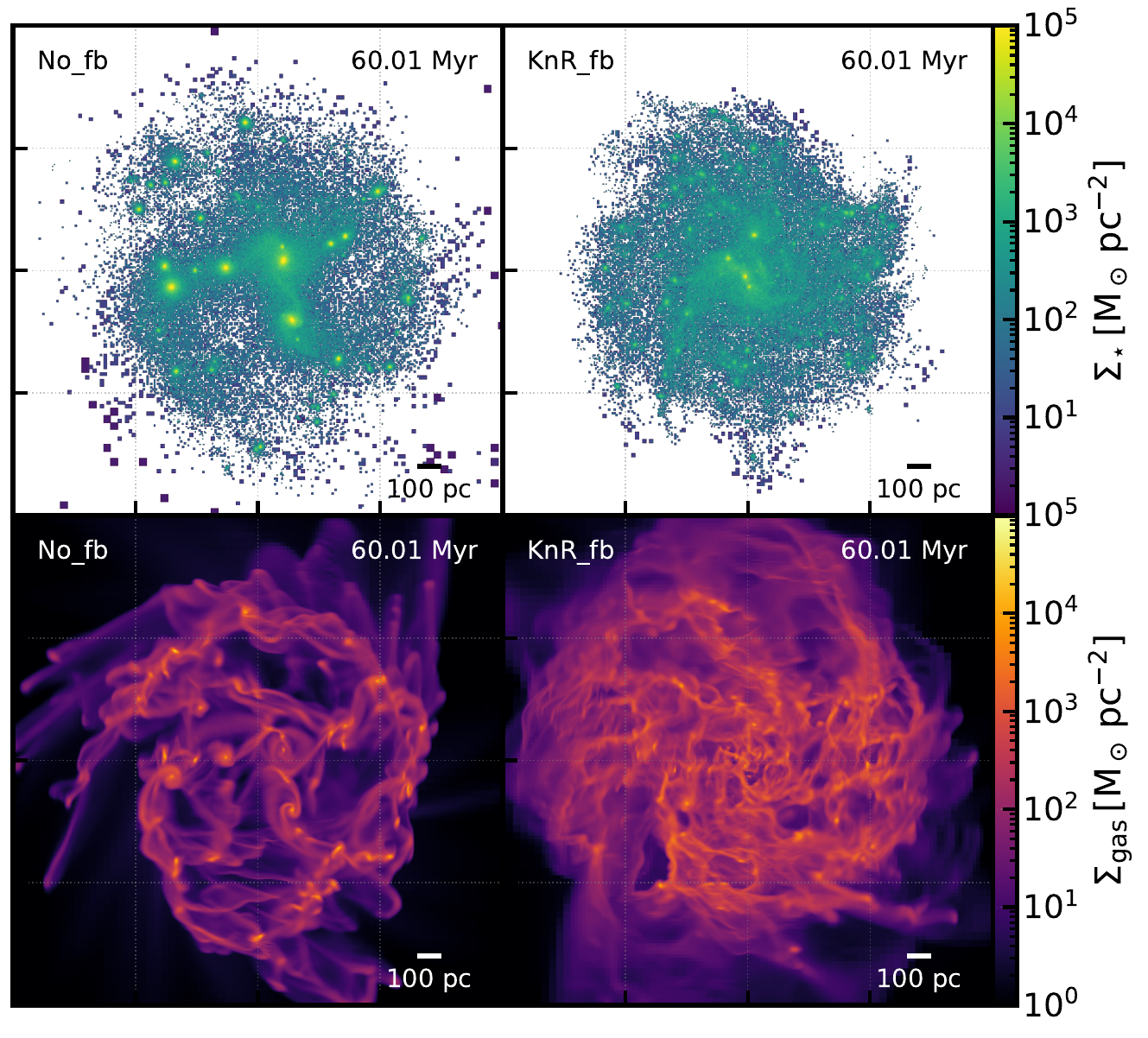}
    \caption{Face-on view of stellar surface density $\Sigma_\star$ (top) and gas column density $\Sigma_{\rm gas}$ (bottom) for \texttt{No\_fb} (left) and \texttt{KnR\_fb} (right) at 60 Myr. The name, time, and scale of the simulation runs are shown in the top left, top right, and bottom right corner of each panel.}
    \label{fig:nofb_vs_kin_rt}
\end{figure}

Although feedback produces only moderate changes in the galaxy-integrated star formation history, it strongly affects the spatial distribution of the gas and its association with stellar clumps. To illustrate this point, Fig.~\ref{fig:nofb_vs_kin_rt} compares the \texttt{No\_fb} and \texttt{KnR\_fb} (the fiducial) runs at $t=60\,{\rm Myr}$. These runs represent the two limiting outcomes among the feedback prescriptions explored here. In the absence of feedback, the stellar surface-density peaks correspond closely to compact, high-density gas structures. Gas therefore remains concentrated around the sites in which the clumps formed and can continue to fuel their subsequent growth. In the \texttt{KnR\_fb} run, by contrast, the gas distribution is smoother and shows a much weaker correspondence with the stellar overdensities. The stellar clumps persist after their residual gas has been partially dispersed, demonstrating that feedback acts more efficiently on the gaseous component than on the already formed stars.

Interestingly, the two runs have comparable instantaneous SFRs at $t\simeq60\,{\rm Myr}$ despite their markedly different gas morphologies. This does not imply that feedback is dynamically unimportant. The global SFR measures the sum of star formation over all cold and dense cells in the galaxy, whereas the maps show whether that gas remains spatially associated with previously formed stellar
clumps. In the fiducial run, star formation can continue in small, short-lived dense structures, filaments, or gas compressed elsewhere
in the disk even after gas has been removed from older clumps. This interpretation is supported by the clump statistical analysis that
will be discussed later. 

In summary, feedback regulates primarily the location, persistence, and integrated efficiency of star formation, rather than enforcing a uniform suppression of its instantaneous global rate.

\subsection{Disk assembly and structural evolution}\label{sec:disk}

\begin{figure}
    \centering
    \includegraphics[width=1\linewidth]{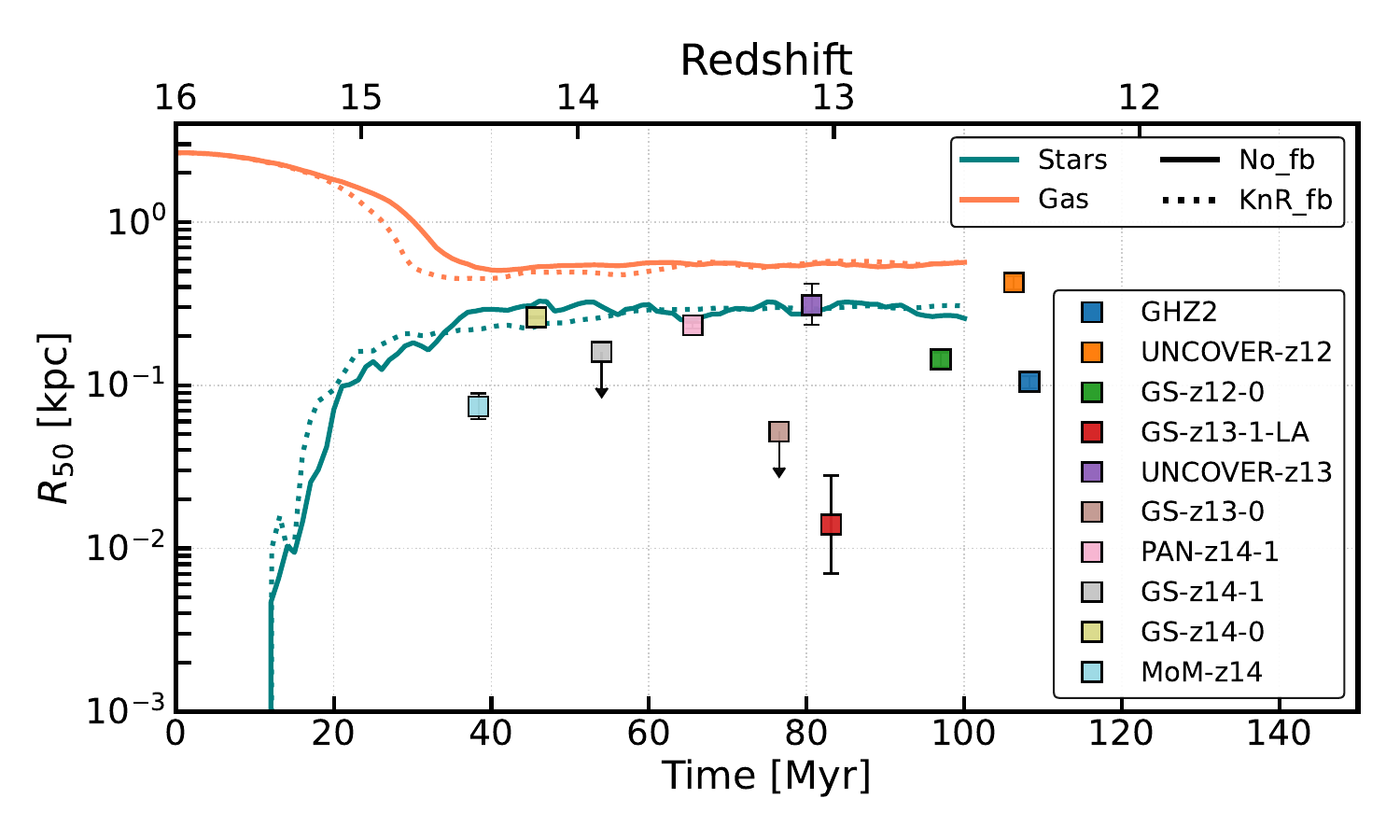}
    \caption{Half-mass radii $R_{50}$ for stars (teal) and gas (coral) as a function of time (lower axis) and redshift (upper axis), measured from the galaxy's center of mass, for \texttt{No\_fb} (solid lines) and \texttt{KnR\_fb} (dotted lines). The squares with error bars show observed high redshift galaxies (see Tab.\ref{tab:observed_galaxies}).}
    \label{fig:halfmassradius}
\end{figure}

The structural evolution of the galaxy is quantified by the half-mass radii in Fig.~\ref{fig:halfmassradius}. The initially extended gas distribution contracts rapidly from $R_{50,\rm gas}\sim3\,{\rm kpc}$ to $\sim0.5\,{\rm kpc}$ during the first $30$--$40\,{\rm Myr}$. At the same time, the stellar half-mass radius grows from a few parsecs immediately after the onset of star formation to $R_{50,\star}\simeq0.2$--$0.3\,{\rm kpc}$. It subsequently remains approximately constant despite continued star formation and feedback. The larger apparent extent of the galaxy in the surface-density maps does not conflict with these values: the maps include diffuse material at large radii, whereas $R_{50}$ measures the radius enclosing half of the mass.

Feedback produces only modest differences in the stellar and gas half-mass radius, despite its much stronger effect on the distribution of dense gas (see Fig. \ref{fig:nofb_vs_kin_rt}). This suggests that feedback primarily regulates the gas retained within individual clumps and the vertical turbulence of the disk, while leaving its characteristic stellar size comparatively unchanged. More generally, the rapid early growth of $R_{50,\star}$ implies that very compact and more extended galaxies at $z>12$ could represent different stages of the same assembly sequence rather than intrinsically distinct populations. 

\begin{figure}
    \centering
    \includegraphics[width=1\linewidth]{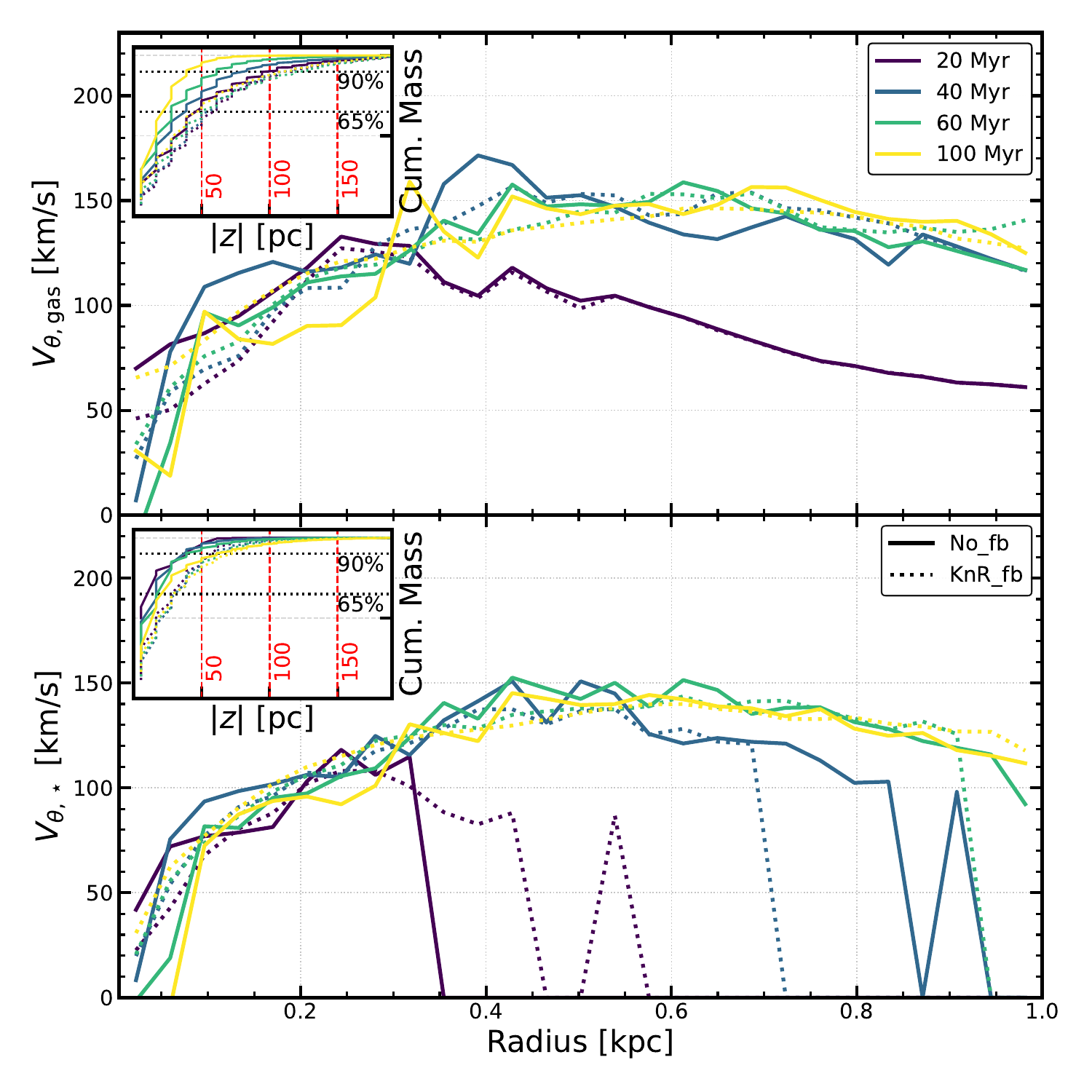}
    \caption{Gas $V_{\theta,\rm gas}$ (top panel) and stellar $V_{\theta,\star}$ (bottom) simulated rotation curves at 20, 40, 60 and 100 Myr for \texttt{No\_fb} (solid lines) and \texttt{KnR\_fb} (dotted lines). Rotation curves are derived from mass‑weighted cylindrical velocity profiles of the galactic disk, centered at the center of mass, assuming a height of 50 pc in either direction of the central plane. The inset shows the cumulative mass fraction of gas above and below the disk midplane (|z|). The gas rotation curve reaches the halo circular velocity (120.7 $\rm km\ s^{-1}$) by 40 Myr.}
    \label{fig:rotationcurve}
\end{figure}

The interpretation of the compact stellar structures as clumps formed in-situ requires the underlying galaxy to develop a rotationally supported disk. We therefore examine the azimuthal kinematics of the gas and stars in Fig.~\ref{fig:rotationcurve}. By $t\simeq20\,{\rm Myr}$, both components exhibit coherent rotation, and by $40\,{\rm Myr}$ the gas reaches rotational velocities of approximately $120$--$150\,{\rm km\,s^{-1}}$ over the inner kiloparsec. These values are comparable to the halo virial velocity, $v_{\rm vir}=120.7\,{\rm km\,s^{-1}}$, showing that the collapsing gas has settled into a dynamically supported disk rather than remaining in radial free fall. The stellar component follows a broadly similar rotation curve, indicating that newly formed stars inherit the angular momentum of the gas.

\begin{figure}
    \centering
    \includegraphics[width=1\linewidth]{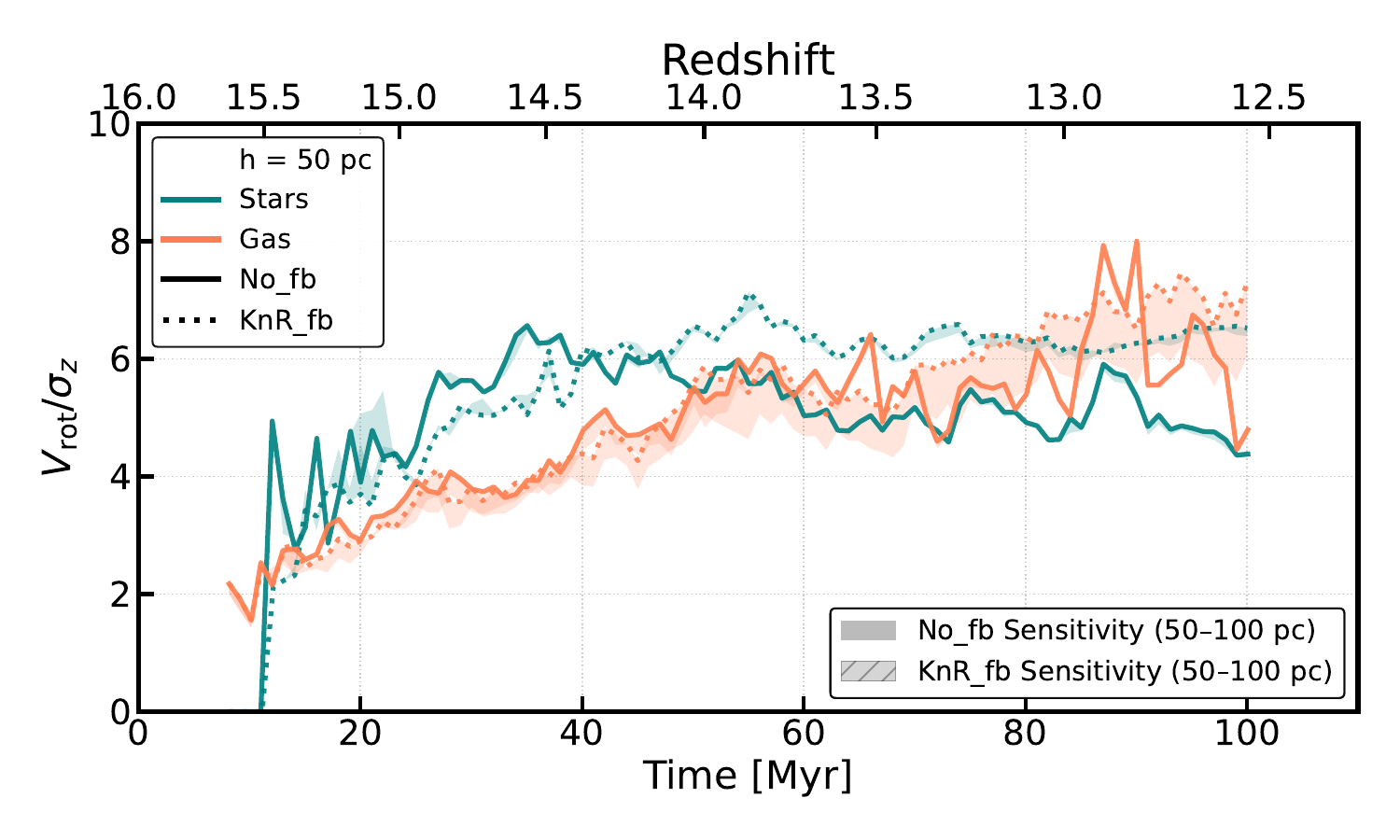}
    \caption{Rotation support $V_{\rm rot} / \sigma_z$ for gas (coral) and stars (teal) as a function of time (lower axis) and redshift (upper axis) for \texttt{No\_fb} (solid lines) and \texttt{KnR\_fb} (dotted lines). Rotation support is quantified as the ratio between the peak azimuthal rotation velocity and the averaged vertical velocity dispersion, i.e. $V_{\theta,\rm max}/\sigma_{z,\rm mean}$, over the disk radius (0.05–0.7 kpc) of thickness 0.05 kpc.}
    \label{fig:rotationsupport}
\end{figure}

The disk remains rotationally supported throughout most of its subsequent evolution. As shown in Fig.~\ref{fig:rotationsupport}, $V_{\theta,\rm max}/\sigma_{z,\rm mean}$, where $V_{\theta,\rm max}$ is the maximum rotation velocity and $\sigma_{z,\rm mean}$ is mean of vertical velocity dispersion over the disk radius (0.05–0.7 kpc) with slab thickness of 50 pc on either side of disk midplane, exceeds unity for both gas and stars after the initial collapse. The stellar component typically maintains $V_{\theta ,\rm max}/\sigma_{z, \rm mean}\simeq4$--$7$, whereas the gas becomes progressively less turbulent. In the no-feedback run, $V_{\theta,\rm max}/\sigma_{z,\rm mean}$ for the gas rises to values of approximately $6$ at late times. Kinetic and radiative feedback instead maintain a larger vertical velocity dispersion. This is evident from the shaded regions; increasing the slab thickness from 50 to 100 pc decreases the rotation support, as the thicker slab incorporates a larger fraction of feedback-deposited energy in gas. Feedback therefore dynamically heats the gas but does not erase the ordered rotation of the disk. This is similar to other analytical model \citep{Krumholz_Burkhart_Forbes_Crocker_2018}, cosmological \citep{Kohandel_Pallottini_Ferrara_Carniani_Gallerani_Vallini_Zanella_Behrens_2020} and isolated \citep{Ejdetjärn_Agertz_Östlin_Renaud_Romeo_2022} simulation studies where gravity and mass transport dominate the disk turbulence, assigning a secondary role to the stellar feedback unlike \cite{Rizzo_Bacchini_Kohandel_Di_Mascolo_Fraternali_Roman-Oliveira_Zanella_Popping_Valentino_Magdis_et_al_2024}. The formation of stellar clumps in our isolated, rotationally supported system is consequently consistent with an in-situ fragmentation pathway, although a direct Toomre analysis is required to establish the specific instability responsible.

\subsection{Onset of gravitational instability}
\label{sec:toomre_results}
\begin{figure}
    \centering
    \includegraphics[width=\linewidth]{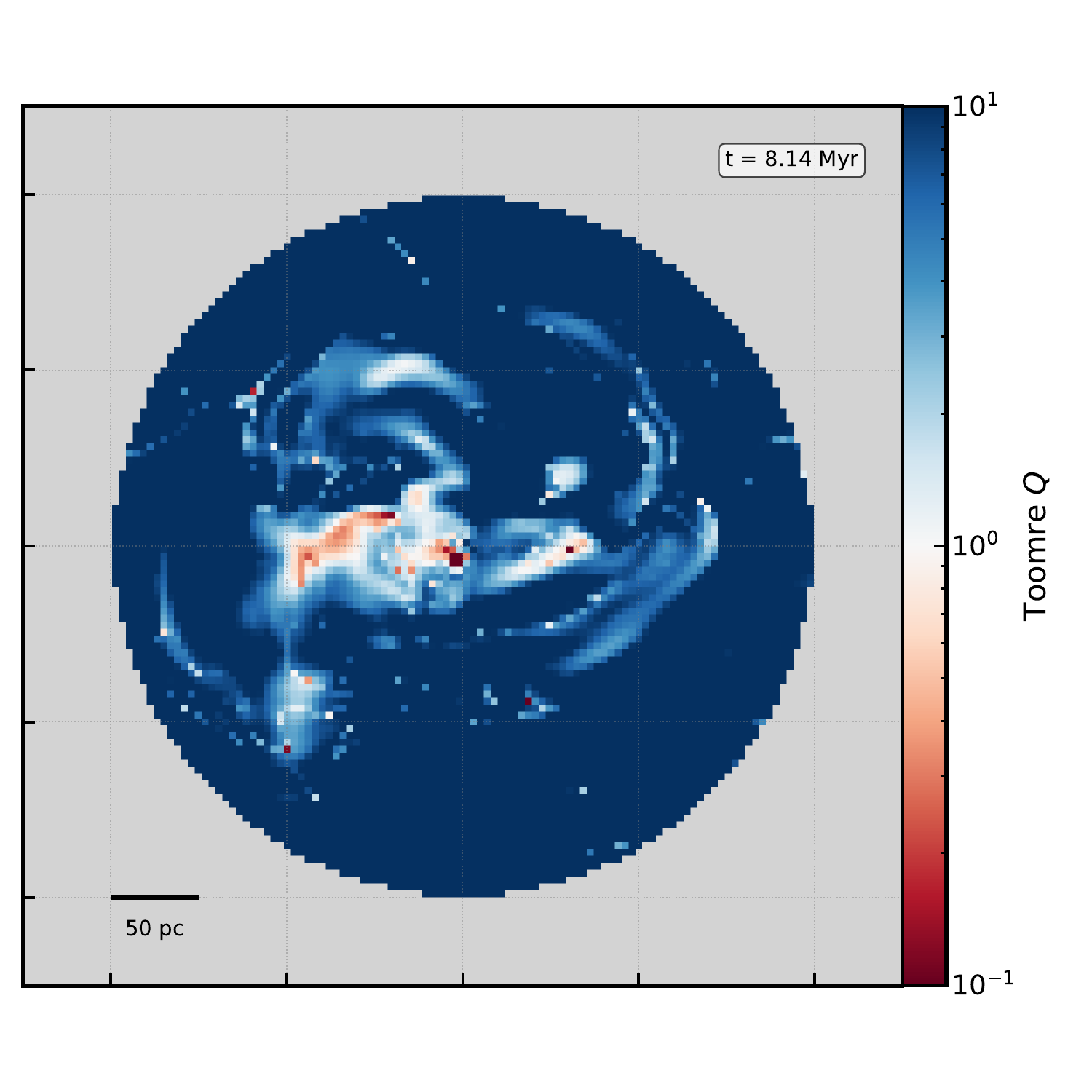}
    \caption{Spatial distribution of the gaseous Toomre parameter $Q_{\rm gas}$ in the \texttt{No\_fb} run at $t=8\,{\rm Myr}$, immediately before the onset of star formation. The map covers the central $400$ pc of the disk. Regions with $Q_{\rm gas}<1$ are unstable in the local thin-disk approximation. The colorbar is capped at $Q_{\rm gas}=10$.}
    \label{fig:Toomremap}
\end{figure}

Having established that the gas settles into a rotationally
supported disk, we investigate whether the onset of clump
formation is preceded by local gravitational instability. We
characterize the disk using the gaseous Toomre parameter,
\begin{equation}
    Q_{\rm gas}(R,\boldsymbol{x}) =
    \frac{c_{\rm eff}(\boldsymbol{x})\kappa(R)}
         {\pi G\Sigma_{\rm gas}(\boldsymbol{x})},
\end{equation}
where $\boldsymbol{x}$ denotes the projected position in the disk
plane and $R_{\boldsymbol{x}}$ is its galactocentric radius, and $c_{\rm eff}$ is the effective gas velocity dispersion,
$\Sigma_{\rm gas}$ is the projected gas surface density, and
$\kappa$ is the epicyclic frequency derived from the azimuthally
averaged rotation curve. Regions with $Q_{\rm gas}<1$ are locally
unstable to axisymmetric perturbations in the thin-disk
approximation. 

For constructing the Toomre Q map, we project the gas within a cylindrical region of radius 0.2 kpc and height 50 pc, on both directions of the central plane, onto a fixed $\approx$ 3.9 pc/pixel grid on a 0.5 kpc field of view centered on Ninfea\_blu. The $c_{\rm eff} = \sqrt{c_s^2 + \sigma^2}$ is computed with $c_s$, local thermal sound speed and $\sigma$, the velocity dispersion derived from the mass weighted variances of the three velocity components in each pixel. We evaluate $\Omega = v_\phi/R$ on the grid and compute
\begin{equation}
\kappa = \sqrt{R\, \partial \Omega^2/\partial R\, +\, 4\Omega^2}
\end{equation}
using finite-difference derivatives of $\Omega^2$ along the x and y directions. \footnote{When $\kappa^2 \leq 0$, we use $\kappa^2 = 2 \times \Omega^2$ to produce a smooth map} Fig.~\ref{fig:Toomremap} shows the spatial distribution of
$Q_{\rm gas}$ in the \texttt{No\_fb} run immediately before the onset of star formation ($t=8\,{\rm Myr}$). Extended
regions of the central disk already have $Q_{\rm gas}<1$, while
no stars have yet formed. Since all simulations share identical initial
conditions and no stellar feedback operates before the first
stars form, this early instability is common to the complete set
of runs.

The temporal ordering where $Q_{\rm gas}<1$ forms just before the first stellar structures at 9 Myr supports a gravitational fragmentation origin for the clumps. They do not, by themselves, demonstrate that the subsequent masses and evolution of the stellar clumps are determined exclusively by
linear Toomre instability. We return to this distinction in
Sec.~\ref{sec:toomre_discussion}.

\section{Clump properties}\label{sec:clumps}

One of the most striking discoveries brought by JWST lensed observations of early galaxies is the presence of numerous, compact and bright stellar clumps. The clumpy nature of these systems might crucially impact their observed properties and evolution, and have implications for chemical abundances and even the origin of massive black hole seeds. We therefore turn to the analysis of simulated clump properties.

\subsection{Clump identification}\label{sec:selectioncriteria}

We identify stellar clumps in face-on maps of the projected stellar surface density constructed at each simulation output. The maps are centered on the galaxy and oriented perpendicular to the angular-momentum vector of the stellar disk. The projection integrates the stellar distribution through a 2 kpc wide region and renders it onto a fixed grid with a resolution of 3.91 pc/pixel, which is slightly higher than the finest resolution of the simulation. We select all pixels satisfying
\begin{subequations}
\begin{equation}
    \Sigma_\star > \Sigma_{\star,\rm th}
\end{equation}
The r.m.s.\ fluctuation of the stellar surface-density maps varies across simulation outputs, ranging from $\sim 250-600\, \Msun \, {\rm pc}^{-2}$. So, we adopt 
a single fixed threshold of 
\begin{equation}
\Sigma_{\star,\rm th} = 2\times10^3\,\Msun\,{\rm pc}^{-2}
\end{equation}
\end{subequations}
which exceeds the r.m.s.\ by approximately three to seven times.

Before identifying individual peaks, we smooth the maps with a Gaussian kernel of width $\sigma=\Delta x$, where $\Delta x$ is the finest spatial resolution. Local maxima are then identified within the regions above $\Sigma_{\star,\rm th}$. Two maxima are treated as distinct clumps only when their projected separation exceeds a minimum value $d_{\rm peak}=11.72 pc$.

Starting from the identified maxima, pixels belonging to each connected region above the surface-density threshold are assigned to the nearest peak. This procedure divides structures containing multiple maxima into separate clump masks while preserving the full area above $\Sigma_{\star,\rm th}$. For each mask, we define the projected effective radius as $R_{\rm eff}=\sqrt{{A_{\rm cl}}/{\pi}}$, where $A_{\rm cl}$ is the total projected area assigned to the clump. We retain only candidates with $R_{\rm eff}\geq2\Delta x$, thereby excluding peaks whose projected extent is comparable to a single resolution element. \footnote{This size criterion does not imply that the internal structure of the smallest retained clumps is fully resolved.}

\begin{figure}
    \centering
    \includegraphics[width=\linewidth]{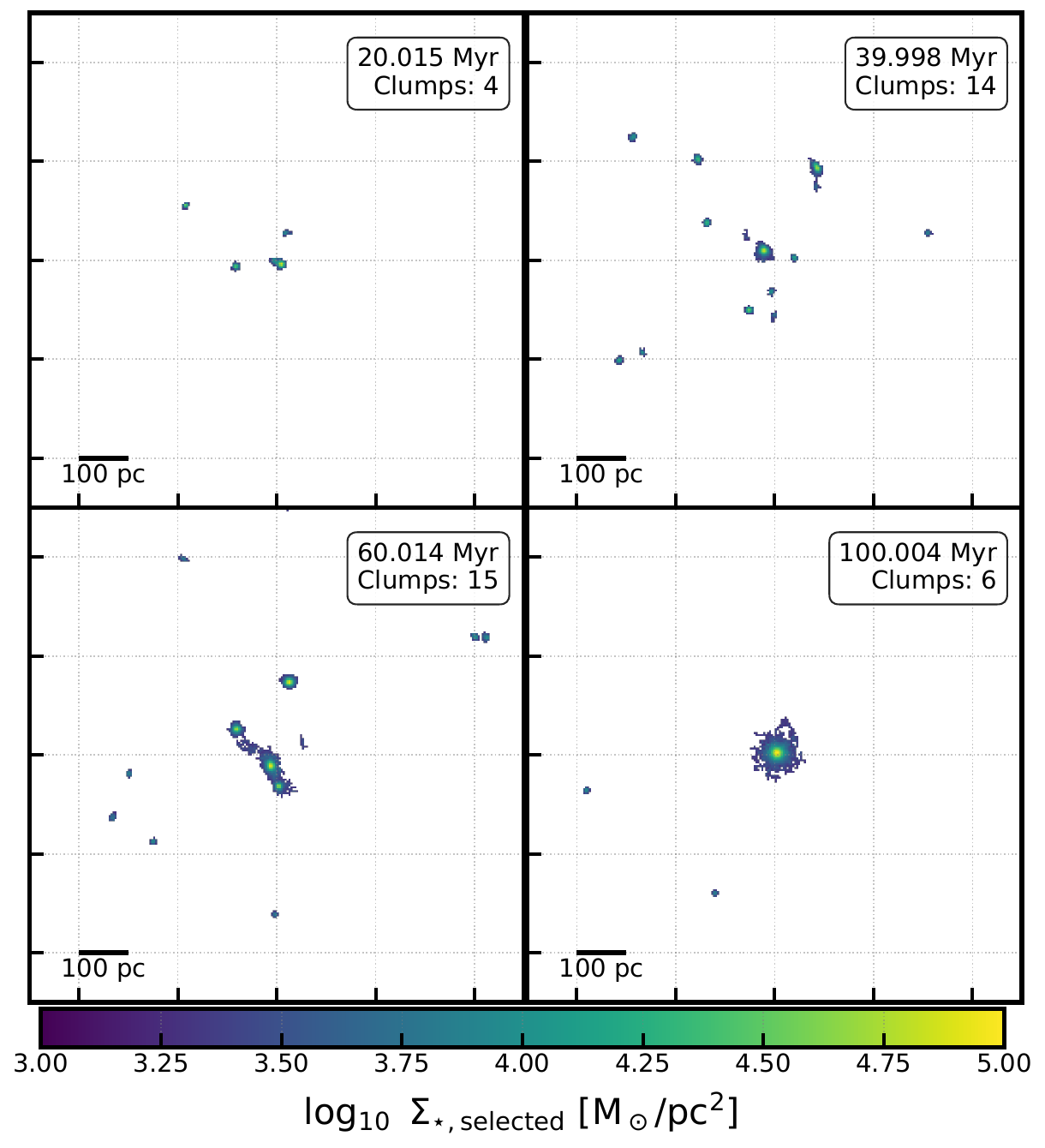}
    \caption{Clumps identified in the stellar surface-density maps
    of the fiducial \texttt{KnR\_fb} run at $t=20$, 40, 60 and
    $100\,{\rm Myr}$. Only pixels assigned to a clump mask are
    shown. The colour scale gives the projected stellar surface
    density, $\Sigma_{\star,\rm selected}$. The simulation time, total number of identified clumps and the scale are reported in each
    panel.}
    \label{fig:Selectedclumps}
\end{figure}

The number and spatial distribution of the selected structures evolve continuously as new clumps form and existing clumps merge, migrate or disperse. Fig.~\ref{fig:Selectedclumps} illustrates the resulting clump catalog for the fiducial \texttt{KnR\_fb} run at $t=20$, 40, 60 and $100\,{\rm Myr}$.  We assess the sensitivity of the catalog to the adopted surface-density threshold and minimum peak separation in Appendix~\ref{app:select}. Although the number of detected low-mass clumps varies with these choices, the
inferred slope of the clump mass function is considerably more stable.

\subsection{Clump population properties}
\label{sec:clump_properties}

For every clump identified as described in
Sec.~\ref{sec:selectioncriteria}, we compute the stellar mass by
integrating the stellar surface density over its projected mask,
\begin{equation}
    M_\star=\int_{A_{\rm cl}}\Sigma_\star\,{\rm d}A.
\end{equation}
The gas mass is obtained by integrating the gas surface density
over the same mask. We stress that this quantity measures the gas
projected within the stellar extent of the clump and does not
necessarily include all gas gravitationally bound to it. Conversely,
it may contain some unrelated gas projected along the line of sight.

\begin{figure}
    \centering
    \includegraphics[width=1\linewidth]{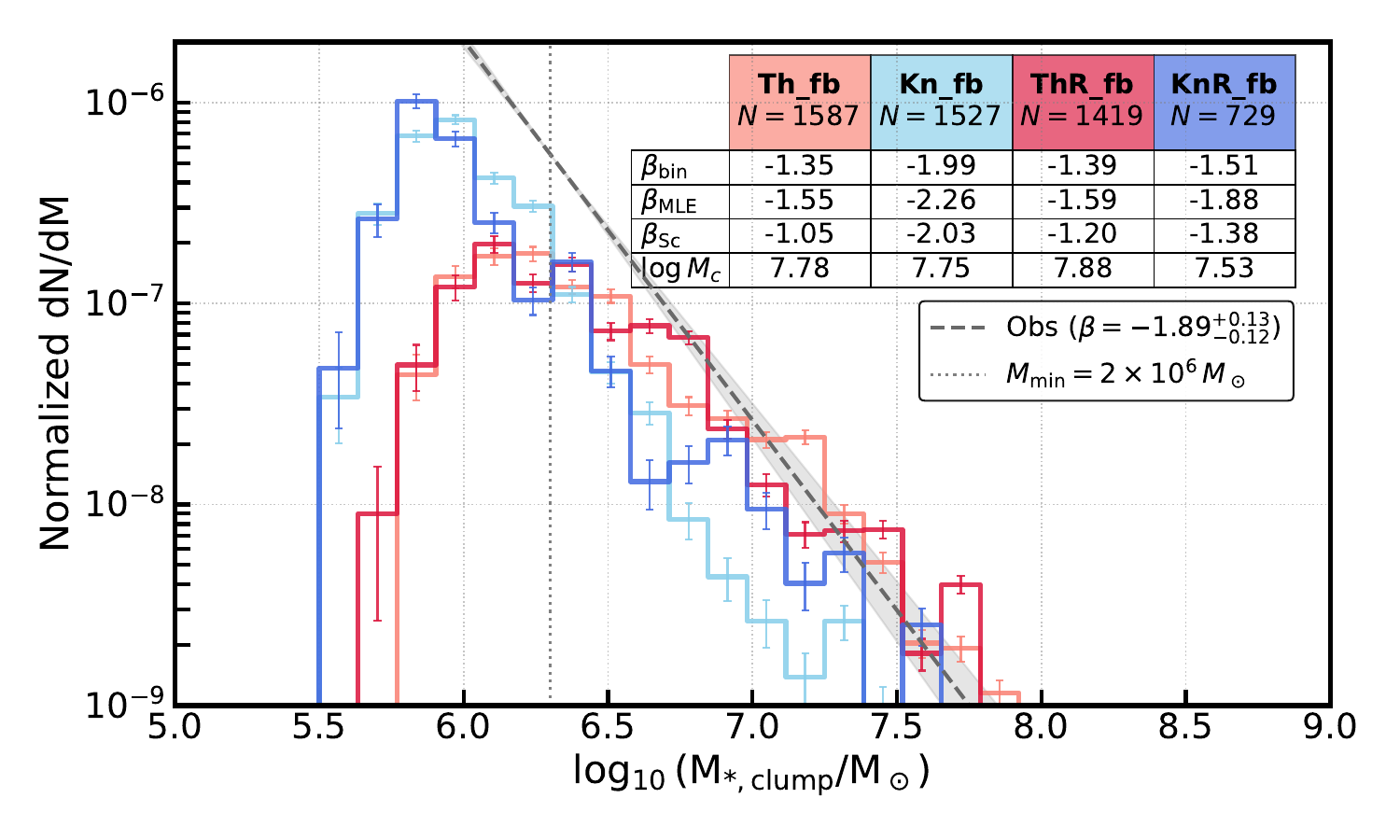}
    \caption{Snapshot-aggregated stellar clump mass functions for \texttt{Th\_fb}, \texttt{ThR\_fb}, \texttt{Kn\_fb} and \texttt{KnR\_fb}. The error bars are the Poisson uncertainty of the counts per bin. The vertical dotted line marks the adopted lower fitting limit, $M_{\rm min}=2\times10^6\,\Msun$. The gray dashed line and shaded region show the observed power-law slope and its uncertainty from \citet{Claeyssens_Adamo_Kokorev_Furtak_Richard_Beauchesne_Dessauges-Zavadsky_Atek_Chisholm_Endsley_et_al_2026}. The legend reports the number of objects and the parameters obtained from the binned, unbinned power-law and Schechter fits.}    
    \label{fig:mstardistclump}
\end{figure}

\subsubsection{Clump mass function}
Fig.~\ref{fig:mstardistclump} shows the stellar clump mass
functions for the four runs including stellar feedback. We restrict
the analysis to $M_\star\geq M_{\rm min}=2\times10^6\,\Msun$.
This limit is comparable to the observational completeness
threshold adopted by
\citet{Claeyssens_Adamo_Kokorev_Furtak_Richard_Beauchesne_Dessauges-Zavadsky_Atek_Chisholm_Endsley_et_al_2026}.
The imposed upper bound, $M_\star<10^9\,\Msun$, lies above the
mass of all relevant objects and therefore has no appreciable
effect on the inferred distributions.

We characterize the CMF using both a power law,
\begin{equation}
    \frac{{\rm d}N}{{\rm d}M_\star}
      = A M_\star^\beta,
\end{equation}
and a Schechter form,
\begin{equation}
    \frac{{\rm d}N}{{\rm d}M_\star}
      = A M_\star^\beta
        \exp\left(-\frac{M_\star}{M_c}\right).
\end{equation}
We estimate the power-law slope, $\beta$, using an unbinned
maximum-likelihood method, thereby avoiding sensitivity to the
choice of histogram bins. Fits to the binned distribution are
reported as a visual and methodological consistency check, while
the Schechter fit tests for a possible suppression of the
high-mass tail. The maximum-likelihood power-law estimate is
adopted as our fiducial measurement. Slopes for all the methods are estimated over the clump mass interval of $2 \times 10^6 \, -\, 10^9 \,\Msun$.

The thermal-feedback runs yield relatively shallow slopes,
$\beta_{\rm MLE}=-1.55$ and $-1.59$ for \texttt{Th\_fb} and
\texttt{ThR\_fb}, respectively, reflecting their larger relative
abundance of massive clumps. The kinetic-feedback runs produce
steeper distributions, closer to the observed
$\beta_{\rm obs}\simeq-1.89$. This indicates that kinetic feedback
preferentially limits the growth or survival of massive clumps.
With the present analysis, these two possibilities cannot be
distinguished: a deficit of massive clumps could result either
from suppressed initial growth or from more rapid subsequent
disruption. Radiative feedback produces a smaller additional
change in the abundance of clumps.

The comparison with the observed CMF should be regarded as
indicative. The simulated catalogue combines clumps identified across 100 Myr of evolution of a single galaxy, Ninfea\_blu, using a $\Delta t=1\, Myr$ time binning, whereas the observational sample contains different galaxies observed at single epochs and is subject to lensing, resolution and completeness effects.
\footnote{Because each snapshot is processed independently, the CMF is a snapshot aggregated distribution. If the same simulated clump
is present in several consecutive outputs, it contributes more
than once and the resulting CMF is weighted by clump lifetime.}

\begin{figure*}
    \centering
    \includegraphics[width=\linewidth]
    {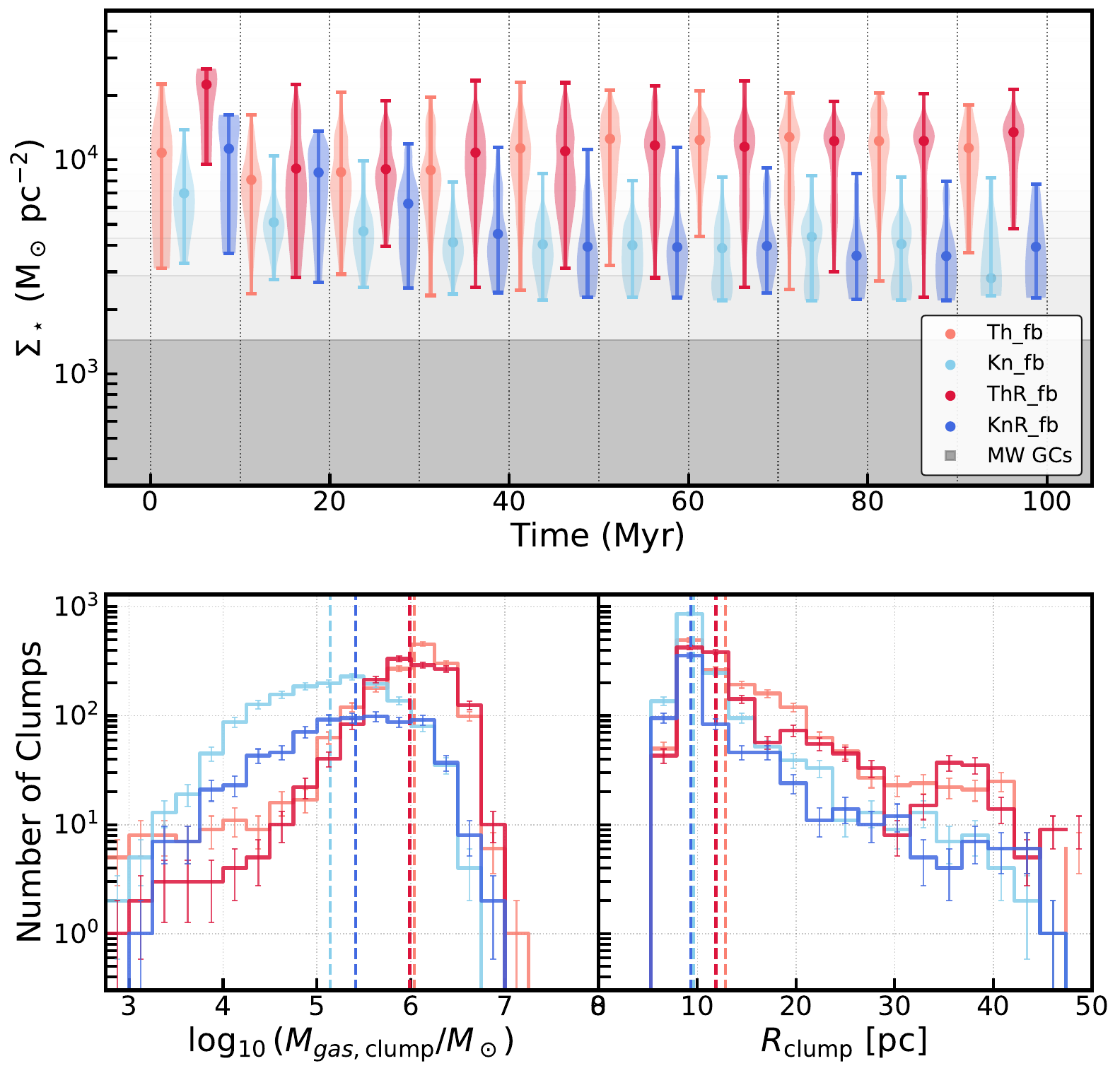}
    \caption{
    Structural properties of the simulated stellar clumps for the
    four feedback runs. \textit{Top}: evolution of the median stellar
    surface density; shaded regions show the corresponding scatter of MW GCs.
    \textit{Bottom left}: distribution of the gas mass projected within the stellar
    clump masks. \textit{Bottom right}: distribution of the projected effective
    radius. Vertical dashed lines in the middle and right panels mark
    the median values for each run.
    The error bars are the Poisson uncertainty of the counts per bin.
    \label{fig:clump_structural_properties}
    }
\end{figure*}

\subsubsection{Sizes, surface densities, and gas content}
The evolution of the clump stellar surface-density distribution is shown in Fig.~\ref{fig:clump_structural_properties}. Clumps in the
thermal-feedback runs maintain characteristic surface densities
of order $10^4\,\Msun\,{\rm pc}^{-2}$, consistent with the weak
regulation already inferred from their CMFs. Kinetic feedback
reduces the median surface density to approximately
$5\times10^3\,\Msun\,{\rm pc}^{-2}$. Thus, the primary effect of
kinetic feedback is not to prevent clump formation, but to limit
the amount of stellar mass accumulated within a given projected
area. By the end of the simulation, the characteristic surface
densities overlap those of the densest Milky Way globular
clusters, although this similarity alone does not establish an
evolutionary connection.

The corresponding gas-mass and radius distributions are shown
in Fig.~\ref{fig:clump_structural_properties}. In the thermal-feedback runs,
the gas projected within stellar-clump masks spans approximately
$10^5$--$10^7\,\Msun$. Kinetic feedback shifts this distribution
towards lower values by about one order of magnitude, confirming
that it efficiently removes gas from the sites of previous star
formation. The effective-radius distributions are considerably
less sensitive to the feedback prescription and peak near
$R_{\rm eff}\sim10$ pc. This peak lies close to the adopted
minimum resolved radius, $R_{\rm eff}=2\Delta x\simeq7.2$ pc,
and must therefore be interpreted cautiously. The simulations
clearly show that compact clumps can form and survive, but do not yet
establish that $10$ pc is a numerically converged characteristic
scale.

\begin{figure}
   \centering
   \includegraphics[width=1\linewidth]{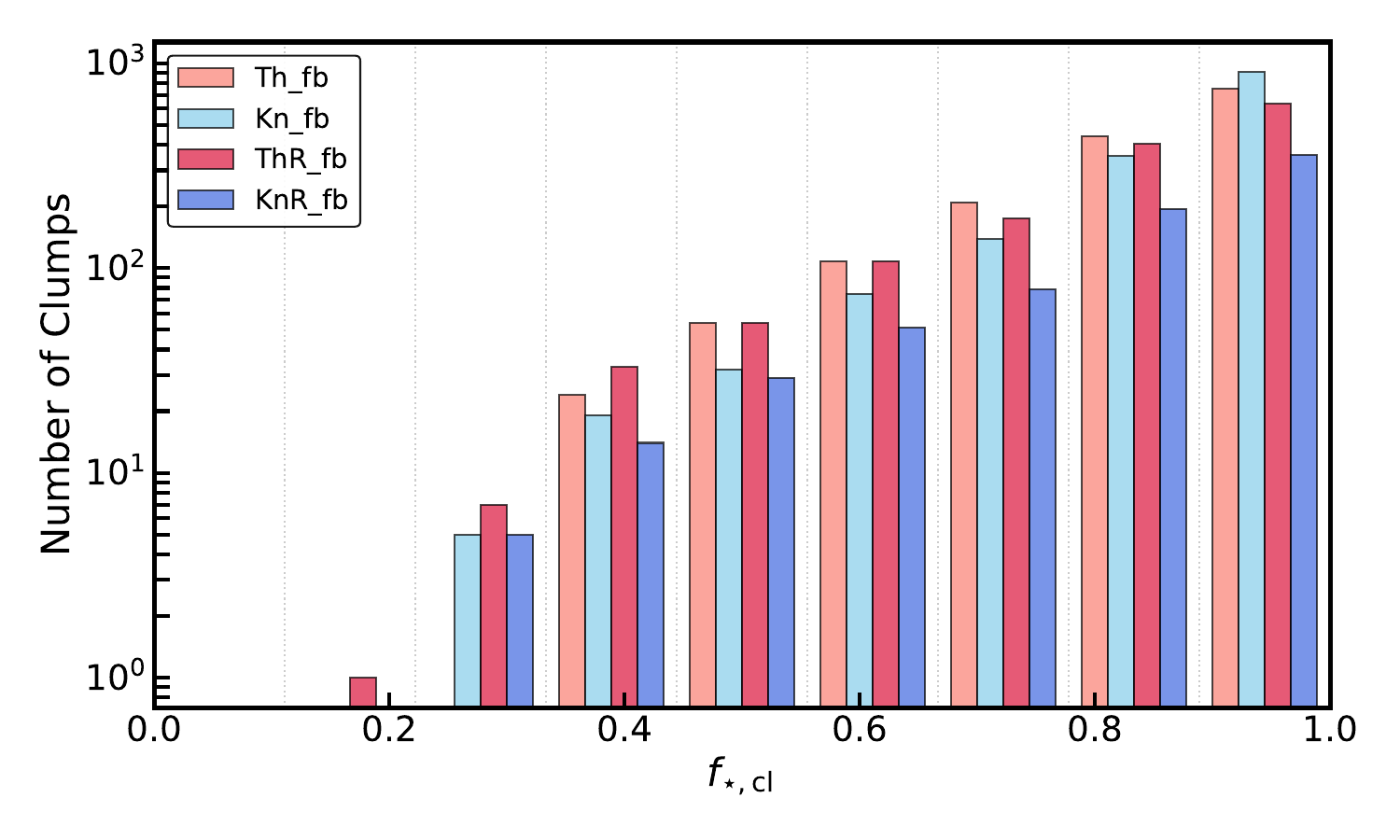}
   \caption{Distribution of the instantaneous stellar mass fraction $f_{\star,\rm cl}=M_\star/(M_\star+M_{\rm gas})$ measured within the projected stellar-clump masks. Colours identify the four feedback runs. The vertical dotted lines mark the histogram-bin boundaries.}  
   \label{fig:clumpefficiency}
\end{figure}

\subsubsection{Stellar and gas fractions}
The instantaneous stellar mass fraction within the projected stellar-clump mask is given by:
\begin{equation}
    f_{\star,\rm cl}=
    \frac{M_\star}{M_\star+M_{\rm gas}}.
\end{equation}
Most selected clumps have $f_{\star,\rm cl}>0.5$ (see Fig. \ref{fig:clumpefficiency}), implying that
their projected baryonic mass is dominated by stars at the time
they are identified. This quantity should not be interpreted as
the integrated star-formation efficiency: it contains no
information about the initial gas mass from which the clump
formed, and it can increase when feedback expels gas from the
clump mask. Its variation among the runs therefore reflects both
the conversion of gas into stars and the subsequent removal or
redistribution of the residual gas.

\subsection{Clump dynamics and migration}\label{sec:clump_dynamics}
\begin{figure}
    \centering
    \includegraphics[width=\linewidth]{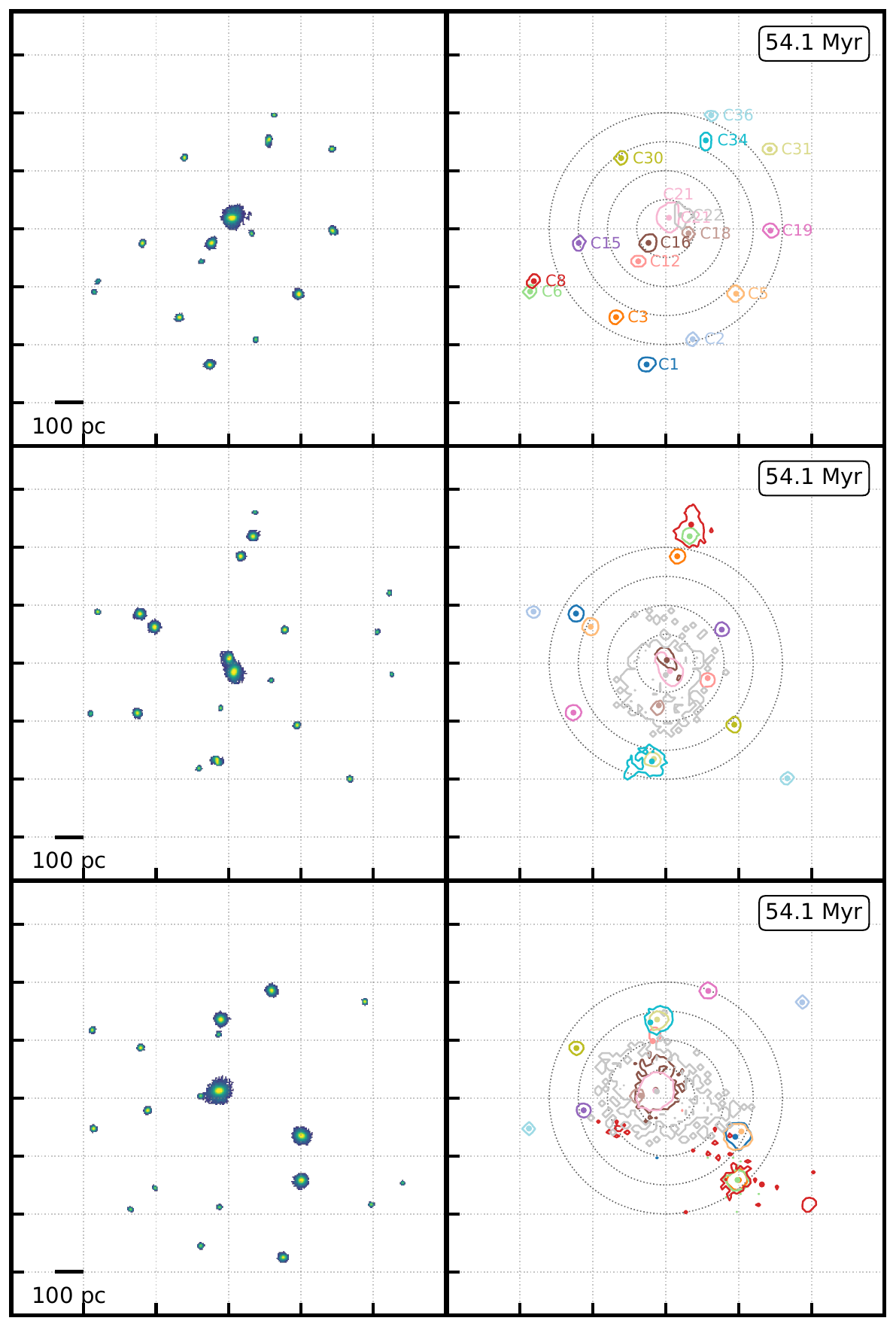}
    \caption{Dynamical evolution of stellar clumps in the
    \texttt{No\_fb} run between $t=40$ and $54\,{\rm Myr}$.
    \textit{Left}: clumps independently identified at $t=40$, 47 and
    $54\,{\rm Myr}$, including only systems with
    $M_\star>2\times10^6\,\Msun$. \textit{Right}: Lagrangian evolution
    of the stellar particles associated with clumps at
    $t=40\,{\rm Myr}$; colours identify their original clump
    membership. Points mark the projected centroids, while contours
    enclose regions with surface density exceeding 0.5 per cent of
    the corresponding peak value to visualize the tidal tails. Black circles mark galactocentric
    radii of 100, 200, 300 and 400 pc.}
    \label{fig:clumpdynamics}
\end{figure}

To investigate the dynamical evolution of individual clumps, we follow the stellar particles associated with clumps identified at $t=40\,{\rm Myr}$ for the subsequent $14\,{\rm Myr}$. We perform this analysis using the \texttt{No\_fb} run in order to isolate evolution driven by gravity, orbital motion, and interactions from the additional effects of stellar feedback. The results therefore provide a baseline for the intrinsic dynamical evolution of the clump population rather than a complete description of the fiducial feedback case.

Fig.~\ref{fig:clumpdynamics} compares two complementary views of the system. The left panels show the clumps independently identified by the clump-finding algorithm at $t=40$, 47 and $54\,{\rm Myr}$. They therefore include newly formed clumps and allow the membership of existing structures to change between outputs. In the right panels, by contrast, the stellar particles belonging to each clump at $40\,{\rm Myr}$ are assigned a unique colour and followed at the later times. These panels provide a Lagrangian view of the redistribution of the original clump members, irrespective of whether they are subsequently assigned to the same clump by the identification algorithm.

The clump population evolves substantially over this relatively short interval. New overdensities appear, some existing systems approach and overlap, and stars initially associated with individual clumps can become distributed over elongated structures. The latter provide evidence for tidal deformation and shear produced by the galactic potential and differential rotation. Thus, even in the absence of stellar feedback, the clumps cannot be regarded as isolated systems with fixed masses and memberships.

The orbital evolution depends strongly on both clump mass and galactocentric position. Among the clumps initially located at $r\gtrsim300\,{\rm pc}$, systems with $M_\star\gtrsim10^7\,\Msun$ show a systematic inward displacement, with radial migration speeds of approximately $15-40\,{\rm km\,s^{-1}}$, inferred from the change in galactocentric radius between consecutive snapshots. Such behaviour is qualitatively consistent with angular-momentum loss through dynamical friction and gravitational torques. Lower-mass clumps show less coherent radial evolution over the same interval, while clumps already in
the central region follow complex orbits and may oscillate about the galactic centre. Given the limited tracking time, these trajectories should not be interpreted as evidence that the lower-mass systems remain permanently on stable orbits.

As massive clumps migrate inward, their projected distributions occasionally converge and their stellar populations overlap, suggesting interactions or coalescence. Other systems become stretched into tidal features and progressively lose their spatial identity. Confirming whether two clumps physically merge, or whether an individual clump becomes gravitationally unbound, would require a particle-based descendant catalogue together with an analysis of the binding energy. We therefore interpret Fig.~\ref{fig:clumpdynamics} as qualitative evidence that migration, interactions, tidal stripping and dispersal all contribute to the evolution of the clump population before the additional action of stellar feedback.

\subsection{Stellar ages and UV contribution}\label{sec:ageUV}

\begin{figure}
    \centering
    \includegraphics[width=\linewidth]{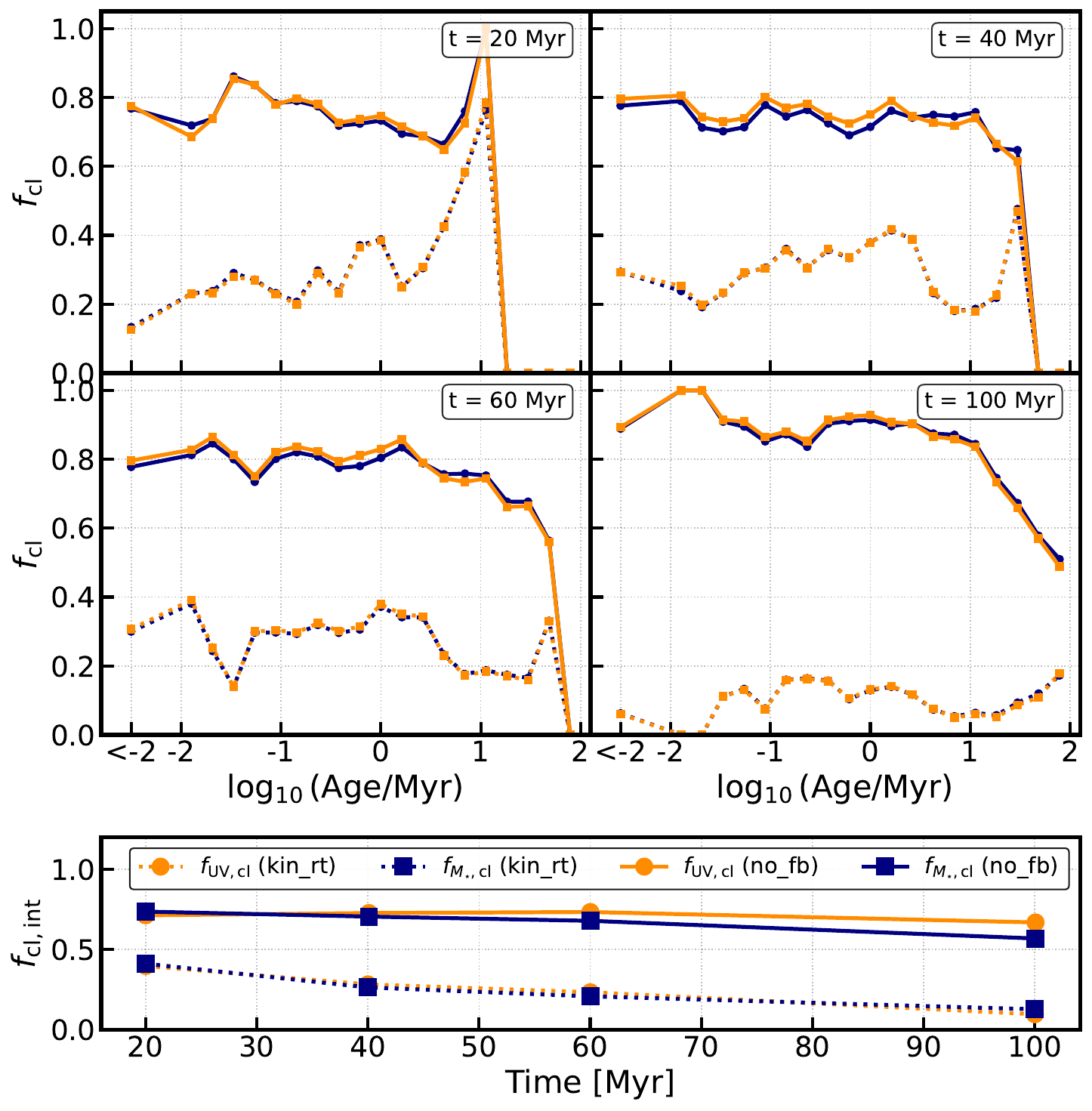}
    \caption{Clump membership fraction for stellar mass $f_{\rm M_{\star}, cl}$ and UV $f_{\rm UV, cl}$ (top four panels) and time evolution $f_{\rm M_{\star}, cl}$ and UV $f_{\rm UV, cl}$ integrated over age (bottom). Each of the top four panels correspond to $t=20\,{\rm Myr}$, $40\,{\rm Myr}$, $60\,{\rm Myr}$ and $100\,{\rm Myr}$ for runs \texttt{No\_fb} and
    \texttt{KnR\_fb}. The clump membership fractions are defined as $ f_{\rm M_{\star}, cl}(t_{\rm age})= M_{\star,\rm cl}(t_{\rm age})/ M_{\star,\rm tot}(t_{\rm age})$ and $ f_{\rm UV, cl}(t_{\rm age})= L_{\rm UV, cl}(t_{\rm age})/ L_{\rm UV, tot}(t_{\rm age}) $. Clump membership is
    assigned from the instantaneous projected position of each
    particle and does not necessarily imply that the particle
    formed in, or remains gravitationally bound to, the clump.
    The histograms show raw particle counts and are
    not stacked.}
    \label{fig:UVmassagecounts}
\end{figure}

Fig.~\ref{fig:UVmassagecounts} presents the clump membership fractions $ f_{\rm M_{\star}, cl}(t_{\rm age})$ ($ f_{\rm UV, cl}(t_{\rm age})$) defined as the ratio of stellar mass (UV luminosity) of stars in the clump to the total stellar mass (UV luminosity) at $t=20$ and $100\,{\rm Myr}$ for the \texttt{No\_fb} and \texttt{KnR\_fb} runs. This classification refers to the instantaneous position of each particle at the time of the snapshot. It therefore measures membership in the identified clump regions, rather than the stellar birth environment or the gravitationally bound fraction of a clump.

In the absence of feedback, stars spanning a broad range of ages remain associated with the clump masks. This indicates that dense stellar structures can retain a substantial fraction of their members, although the presence of stars outside the masks shows that orbital evolution, tidal stripping and shear redistribute some of the stellar population even in the \texttt{No\_fb} run. In the \texttt{KnR\_fb} run, the diffuse component is more prominent over almost the entire age range. Feedback therefore reduces clump membership without producing an obvious strong preference for a particular stellar age.

The age distributions alone cannot determine whether diffuse stars formed outside clumps or were removed from them after formation. Both processes may contribute. Feedback can disperse the natal gas, inhibit the subsequent growth of a stellar overdensity and reduce its gravitational binding, thereby making its stars more susceptible to tidal stripping and dynamical dispersal. Establishing the relative importance of in-situ diffuse star formation and stellar escape would require tracking the birth positions and subsequent trajectories of individual stellar particles.

We next quantify the contribution of the identified clumps to the intrinsic rest-frame UV emission at $1500\,\AA$. Each stellar particle is treated as a simple stellar population, and its UV luminosity is obtained by interpolating the \citet{2003MNRAS.344.1000B} spectral-synthesis tables at the particle age and metallicity. As the BC03 library does not include tracks for metal-free (Z < $10^{-4}$) stars (and Z > $5 \times 10^{-2}$), we clamp our stellar particles, falling outside the range, to the nearest available boundaries. We define $L_{1500}$ as the integrated UV luminosity over $1300$--$1700\,\AA$. The values quoted below are intrinsic and do not include attenuation by dust.

For the \texttt{No\_fb} (\texttt{KnR\_fb}) run, the fraction of the total UV luminosity produced within clump masks is $f_{\rm UV,cl}= [0.71,\ 0.73,\ 0.73, 0.67] \,([0.40,\ 0.28,\ 0.23, 0.10])$ at $t=20$, 40, 60 and $100\,{\rm Myr}$, respectively. The corresponding fractions of stellar mass are
$f_{\star,\rm cl}= [0.74,\ 0.70,\ 0.68,\  0.57]\, ([0.41,\ 0.26,\ 0.21,\  0.13])$. Clumps therefore provide a substantial, and (but not) dominant, fraction of the intrinsic UV emission at early times, and their contribution declines as the galaxy evolves.

The close correspondence between $f_{\rm UV,cl}$ and $f_{\star,\rm cl}$ implies that the mean UV luminosity per unit stellar mass is similar inside and outside the selected clumps. The declining UV fraction is consequently driven mainly by the decreasing fraction of stellar mass associated with clump masks, rather than by differential UV fading alone. This evolution is consistent with a combination of clump dispersal, stellar stripping and continued star formation in the diffuse component. The present analysis does not by itself distinguish among these processes.

\subsection{Comparison with observed stellar systems}\label{sec:compare}

\begin{figure*}
    \centering
    \includegraphics[width=1\linewidth]{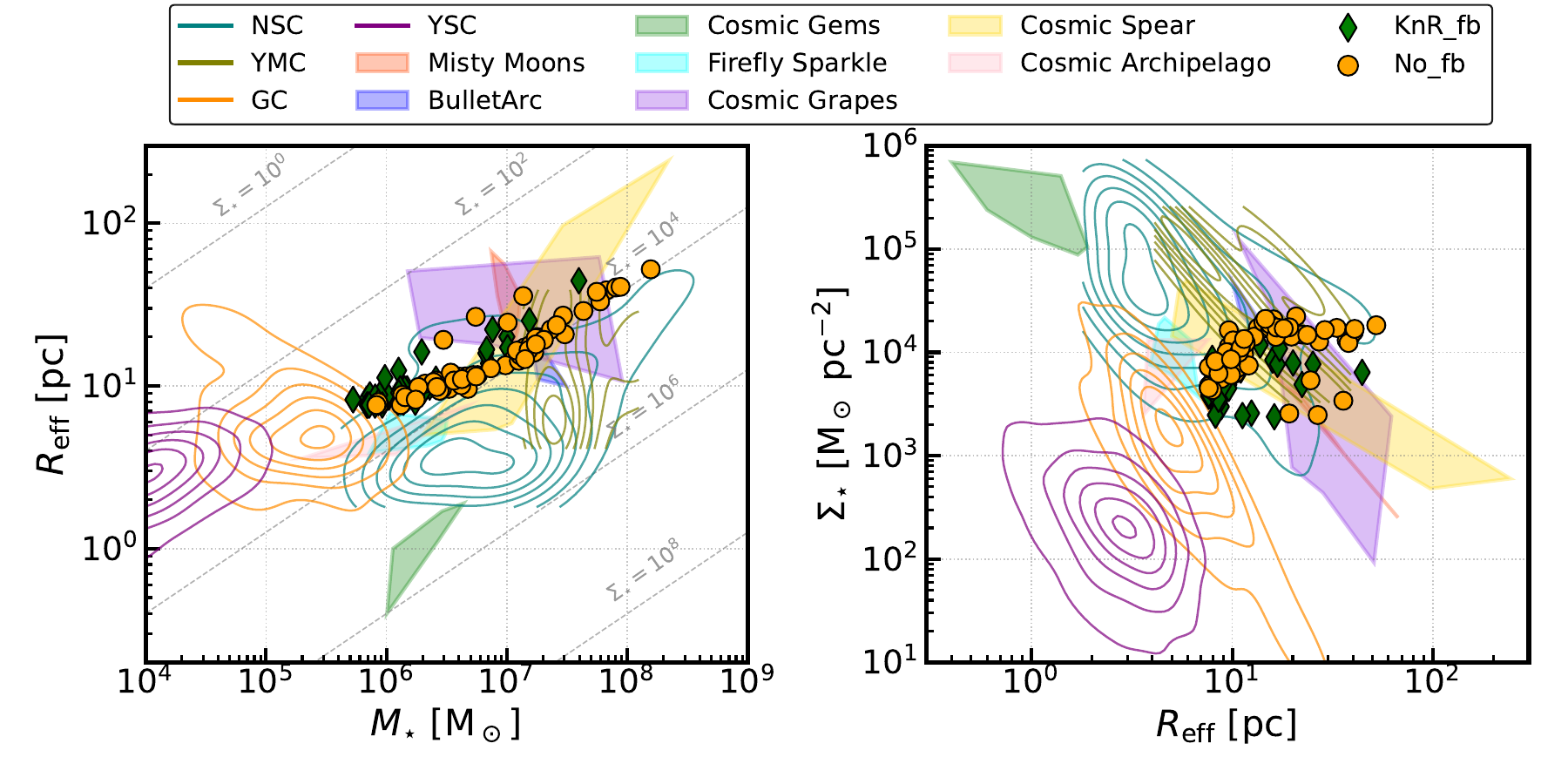}
    \caption{Comparison between simulated clumps and observed compact stellar systems. \textit{Left}: stellar mass versus projected effective radius. \textit{Right}: mean stellar surface density versus effective radius. Orange circles and green diamonds show
    clumps from the \texttt{No\_fb} and \texttt{KnR\_fb} runs, respectively, combining the outputs at $t=20$, 40, 60 and $100\,{\rm Myr}$. The contours represent local star clusters: NSCs (teal) and YMCs (olive) from \cite{Norris_Kannappan_Forbes_Romanowsky_Brodie_Faifer_Huxor_Maraston_Moffett_Penny_et_al._2014}, GCs (darkorange) from \cite{2018MNRAS.478.1520B} and YSCs (purple) from \cite{Brown_Gnedin_2021}. The $z>6$ systems are plotted as convex hulls: Misty Moons (red, \citealt{Nakane_Kokorev_Fujimoto_Ouchi_McLeod_Golubchik_Oguri_Zitrin_Bondestam_Donnan_et_al_2025}), BulletArc-z11 (blue, \citealt{Bradac_Judez_Willott_Rihtarsic_Martis_Harshan_Felicioni_Asada_Desprez_Clowe_et_al_2025}), Cosmic Gems (green, \citealt{Adamo_Bradley_Vanzella_Claeyssens_Welch_Diego_Mahler_Oguri_Sharon_Abdurrouf_et_al_2024}), Firefly Sparkle (cyan, \citealt{Mowla_Iyer_Asada_Desprez_Tan_Martis_Sarrouh_Strait_Abraham_Bradač_et_al_2024}), Cosmic Grapes (blueviolet, \citealt{Fujimoto_Ouchi_Kohno_Valentino_Gimenez-Arteaga_Brammer_Furtak_Kohandel_Oguri_Pallottini_et_al_2025}), Cosmic Spear (gold, \citealt{Abdurrouf_Coe_Resseguier_Murphy_Xu_Adamo_Roy_Henry_Kokorev_Brammer_et_al._2025}) and Cosmic Archipelago (lightpink, \citealt{Messa_Vanzella_Loiacono_Bergamini_Castellano_Sun_Willott_Windhorst_Yan_Angora_et_al_2025}). Dashed gray lines on the left panel show lines of constant stellar surface density in units of $M_\odot\ \rm pc^{-2}$.}
    \label{fig:observedsigmaMrclump}
\end{figure*}

Fig.~\ref{fig:observedsigmaMrclump} compares the simulated clumps with compact stellar systems observed both at high redshift
and in the local Universe. The high-redshift compilation includes the Misty Moons
\citep{Nakane_Kokorev_Fujimoto_Ouchi_McLeod_Golubchik_Oguri_Zitrin_Bondestam_Donnan_et_al_2025},
BulletArc-z11
\citep{Bradac_Judez_Willott_Rihtarsic_Martis_Harshan_Felicioni_Asada_Desprez_Clowe_et_al_2025},
the Cosmic Gems
\citep{Adamo_Bradley_Vanzella_Claeyssens_Welch_Diego_Mahler_Oguri_Sharon_Abdurrouf_et_al_2024},
the Firefly Sparkle
\citep{Mowla_Iyer_Asada_Desprez_Tan_Martis_Sarrouh_Strait_Abraham_Bradač_et_al_2024},
the Cosmic Archipelago
\citep{Messa_Vanzella_Loiacono_Bergamini_Castellano_Sun_Willott_Windhorst_Yan_Angora_et_al_2025},
the Cosmic Grapes
\citep{Fujimoto_Ouchi_Kohno_Valentino_Gimenez-Arteaga_Brammer_Furtak_Kohandel_Oguri_Pallottini_et_al_2025},
and the Cosmic Spear
\citep{Abdurrouf_Coe_Resseguier_Murphy_Xu_Adamo_Roy_Henry_Kokorev_Brammer_et_al._2025}.
For context, we also show the distributions of local nuclear star clusters and young massive clusters
\citep{Norris_Kannappan_Forbes_Romanowsky_Brodie_Faifer_Huxor_Maraston_Moffett_Penny_et_al._2014},
Milky Way globular clusters
\citep{2018MNRAS.478.1520B}, and young star clusters
\citep{Brown_Gnedin_2021}.

The simulated clumps span approximately $M_\star\simeq10^6$--$2\times10^8\,\Msun$ and $R_{\rm eff}\simeq7$--$50\,{\rm pc}$. They overlap a substantial fraction of the region occupied by the observed $z>6$ systems, showing that in-situ fragmentation of a compact, rotationally
supported disk can produce stellar structures with realistic masses and sizes. The agreement is particularly notable because the clumps form self-consistently and are not inserted into the initial conditions.

The most striking effect of feedback is to displace the clumps approximately along loci of constant stellar surface density in the mass-radius plane.  In the \texttt{No\_fb} run, clumps extend to larger masses and radii, whereas \texttt{KnR\_fb} preferentially populates the lower-mass and smaller-radius region. The displacement occurs approximately along loci of constant stellar surface density: both runs produce clumps primarily within
\begin{equation}
  \overline{\Sigma}_\star\simeq
  2\times10^3-3\times10^4\,
  \Msun\,{\rm pc}^{-2}.
\end{equation}
This suggests that feedback principally limits the continued growth of clumps and the area over which they remain coherent, rather than changing their characteristic stellar surface density by a comparable factor. The lower edge of this interval must, however, be interpreted cautiously because the clump catalogue is constructed using $\Sigma_{\star,\rm th}=2\times10^3\,\Msun\,{\rm pc}^{-2}$.

The simulations do not cover the full range of observed properties. In particular, some of the Cosmic Gems have $R_{\rm eff}\lesssim {\rm few}\,{\rm pc}$ and $\overline{\Sigma}_\star\gtrsim10^5\,\Msun\,{\rm pc}^{-2}$, placing them outside the region reached by our clumps. Such
objects are smaller than, or comparable to, the effective resolution limit of the present analysis and therefore cannot be used as a stringent test of the model. Higher-resolution simulations would be required to establish whether the same fragmentation process can produce this extreme population.

The simulated clumps also overlap the high-mass and high-surface-density tails of local globular clusters, young massive clusters and nuclear star clusters. This overlap is suggestive, but it does not establish that the simulated systems are their direct progenitors. Over subsequent cosmic time, stellar evolution, gas loss, tidal stripping, dynamical heating and evaporation can substantially modify both their masses and radii. The comparison therefore demonstrates similarity at formation, whereas any evolutionary connection with present-day compact stellar systems requires modeling their long-term survival.

\section{Discussion}\label{sec:discussion}

\subsection{The origin of the clumps}\label{sec:toomre_discussion}

The controlled nature of the simulations allows us to exclude
mergers, accreted satellites, and externally induced tidal
perturbations as the origin of the first clumps. The disk is rotationally supported before fragmentation, and the first stellar structures form after the disk forms $Q_{\rm gas}<1$ regions in the preceding output. 

The classical Toomre criterion nevertheless provides only an approximate description of the simulated system. It assumes a stationary, infinitesimally thin, axisymmetric and single-component disk (see, e.g. \citealt{2013MNRAS.433.1389R}), whereas the simulated galaxy is rapidly assembling, turbulent, multiphase and has a finite vertical thickness. After star formation begins, the stellar and gaseous
components are also gravitationally coupled, requiring an effective multi-component stability parameter rather than independent $Q_{\rm gas}$ and $Q_\star$ criteria. Non-axisymmetric modes, turbulent density fluctuations and subsequent clump-clump interactions may additionally affect the fragmentation process.

We therefore interpret the $Q$ maps as evidence for a Toomre-like gravitational fragmentation pathway, but not as proof that the complete clump population is determined by linear Toomre theory. The initial instability may set the characteristic birth scale, while accretion, merging, migration, tidal stripping and stellar feedback subsequently reshape the stellar clump mass function. This distinction is particularly important when comparing an analytical birth spectrum with the snapshot-aggregated stellar CMF measured in the simulations.

\subsection{Feedback regulation and the clump lifecycle}\label{sec:feedback_discussion}

Our results distinguish between three stages of clump evolution:
formation, growth, and survival. Dense clumps form in every run,
indicating that the feedback prescriptions considered here do not
prevent the initial fragmentation of the disk. Feedback becomes
more important after formation, when it regulates the gas retained
by the clumps, their subsequent stellar-mass growth and their
susceptibility to disruption. This explains why feedback can
strongly modify the clump population while producing only moderate
changes in the instantaneous galaxy-integrated SFR: gas removed
from established clumps can be replaced as a source of star
formation by newly formed dense structures elsewhere in the disk.

The contrast between thermal and kinetic supernova feedback
primarily demonstrates sensitivity to the numerical treatment of
unresolved remnants. Thermal energy deposited in dense gas is
rapidly radiated away, whereas kinetic injection couples more
effectively to the surrounding medium and limits clump growth.
These models should therefore not be interpreted as rigorous
physical bounds: the thermal and kinetic phases are parts of the
same supernova-remnant evolution in nature. Similarly, the
relatively modest effect of radiative feedback in our simulations
does not include momentum transfer from resonantly scattered
Ly$\alpha$ photons, which may be important in metal-poor
environments \citep{Ferrara25_a,Manzoni25,Nebrin25}.

Clump evolution is governed by the gravitational processes present in the \texttt{No\_fb} run. Torques and dynamical friction
drive the inward migration of some massive clumps, while tidal
forces, differential rotation and clump--clump interactions
produce stripping, deformation and occasional coalescence.
Dynamical friction does not directly disperse a clump, but can
carry it into central regions where stronger tides and interactions
accelerate its evolution. By removing gas and reducing the depth
of the clump potential, stellar feedback makes these gravitational
processes more effective. The simulated galaxy therefore contains
a continuously evolving population in which clumps form, grow,
migrate, interact and disperse, rather than a fixed collection of
permanently bound stellar systems.

In the \texttt{KnR\_fb} run, the spatial connection between clumps and outflows becomes very evident. Clumps with $M_\star>2\times 10^6 \, \Msun$ can be associated with outflows reaching 50-200 km/s. However, clumps producing weaker outflows, when aggregate, often produce larger collimated outflows at higher altitude above the disk. This behaviour suggests that feedback-driven outflows can emerge from both individual and collective action of multiple dense stellar regions. However, this analysis has been done considering all clumps reside in the midplane of the galaxy. A complete 3D, time dependent, analysis across the various feedback runs will be explored in future work.

\subsection{Observational implications and possible descendants}\label{sec:obs_implications}

The broad agreement between the simulated and observed
mass--radius distributions supports in-situ disk fragmentation
as a viable origin for many of the compact stellar structures
observed at $z>6$. Feedback displaces the simulated clumps
approximately along loci of constant stellar surface density,
suggesting that it primarily limits their growth and coherent
extent rather than setting a substantially different density
scale. The most compact systems, particularly the Cosmic Gems,
remain below the effective resolution of our analysis and cannot
yet provide a stringent test of this scenario.

At $z>12$, structures comparable to the simulated clumps would
generally remain unresolved without strong gravitational
lensing. Apparently compact galaxies may therefore contain
multiple stellar clumps embedded in a more diffuse disk. Their
observed morphology may also depend on evolutionary stage:
our galaxy grows rapidly from a compact to a more extended
stellar configuration without a comparably large change in its
SFR. This provides a possible explanation for the diversity of
sizes inferred among super-early galaxies, although current
stellar-age estimates are too heterogeneous and uncertain to
establish such an evolutionary sequence.

The overlap between simulated clumps and the high-density portion
of the local globular-cluster population makes an evolutionary
connection plausible but not demonstrated. Survival to $z=0$
depends on processes not followed here, including stellar mass
loss, two-body relaxation, evaporation, tidal shocks and the
evolving galactic potential. Our simulations therefore identify
possible initial conditions for compact stellar-cluster
progenitors, rather than predicting their present-day descendants.

\subsection{Limitations of the present study}

The controlled nature of our numerical experiment is both its
principal strength and an important limitation. By simulating an
isolated halo, we can exclude mergers, accreted satellites, and
externally induced tidal perturbations as the origin of the first
clumps. However, the galaxy is not supplied by cosmological gas
accretion and does not experience the evolving environment of a
real high-redshift halo. The decline of the SFR at late times is
therefore affected by the exhaustion, ejection and redistribution
of a finite initial gas reservoir. In a cosmological simulation,
continued inflow could replenish the disk, sustain star formation
and repeatedly modify its stability.

We have considered a single halo mass, concentration, spin
parameter and initial angular-momentum distribution. The
resulting disk and clump properties cannot therefore be taken as
representative of the complete high-redshift galaxy population.
In particular, the fragmentation scale is expected to depend on
halo mass, disk surface density, gas fraction, turbulent velocity
dispersion and angular momentum. A broader parameter survey is
required to determine whether the trends found here persist
across the range of systems observed at $z>6$.

The maximum spatial resolution of $3.6\,{\rm pc}$ allows us to
identify parsec-scale stellar structures but does not fully
resolve their internal dynamics. The recovered radius
distribution peaks near $R_{\rm eff}\sim10\,{\rm pc}$, only
moderately above the adopted minimum size
$R_{\rm eff}=2\Delta x\simeq7.2\,{\rm pc}$. Consequently, the
position and narrowness of this peak may be influenced by
resolution and selection. The simulations cannot address the
internal structure of the most compact observed clumps or
establish whether they fragment into smaller stellar systems.

Clumps are identified in two-dimensional stellar
surface-density maps. This procedure resembles the way observed
clumps are selected and facilitates a direct comparison, but
projected masks do not uniquely define gravitationally bound
three-dimensional systems. Gas unrelated to a clump may be
included along the line of sight, while bound material outside
the projected surface-density threshold may be excluded. In
addition, quantities such as
$M_\star/(M_\star+M_{\rm gas})$ describe the instantaneous
contents of the projected mask and should not be interpreted as
integrated star-formation efficiencies.

The statistical clump catalogue combines multiple simulation
outputs. Unless unique descendants are identified, a clump that
persists across several snapshots enters the catalogue more than
once. The resulting mass and structural distributions are
therefore snapshot-aggregated and give greater statistical weight
to long-lived systems. They should not be interpreted directly
as clump birth functions. Particle-based merger trees will be
required to separate the initial mass spectrum from subsequent
growth, merging and disruption and to measure a genuine clump
lifetime distribution.

The feedback results also depend on the adopted numerical
implementations. Thermal energy deposited in dense cells is
subject to rapid numerical cooling, also because of the adopted resolution, while kinetic injection
bypasses part of the unresolved supernova-remnant evolution, but is dependent on the implementation.
Neither prescription alone provides a complete representation of
physical supernova feedback. The radiative-transfer calculations
include photoionization, photo-heating and direct radiation
pressure but omit resonant Ly$\alpha$ pressure and
infrared multi-scattering. The predicted gas
content, surface density and survival of clumps should therefore
be regarded as model dependent.

Finally, the classical Toomre parameter provides an approximate
diagnostic of gravitational instability. The simulated disk is
rapidly assembling, turbulent, multiphase and of finite thickness,
whereas the standard criterion assumes a stationary,
infinitesimally thin, axisymmetric and single-component disk.
After stars form, a coupled gas--stellar stability analysis is
more appropriate than either $Q_{\rm gas}$ or $Q_\star$ alone.
The spatial and temporal association of regions with
$Q_{\rm gas}<1$ and the first stellar clumps supports a
Toomre-like fragmentation pathway, but does not exclude
turbulent, non-axisymmetric or multi-component effects. A
comparison between the predicted instability scale and the
initial masses and separations of newly formed clumps will be
needed for a more stringent test.

\section{Summary}\label{sec:summary}

JWST observations have revealed compact stellar clumps in galaxies
at $z>6$, but their formation mechanism and response to stellar
feedback remain uncertain. We have investigated whether such
systems can form in-situ through the fragmentation of an early
galactic disk, independently of mergers, satellite accretion and
external tidal perturbations.

Using RAMSES-RT, we performed five simulations of an
isolated galaxy, Ninfea\_blu, with halo mass of $1.5\times10^{10}\,\Msun$, evolved for $100\,{\rm
Myr}$ from $z=16$ to $z\simeq12$ with a maximum spatial resolution
of $3.6\,{\rm pc}$. The runs explore thermal and kinetic supernova
feedback, photoionization, photoheating and direct radiation
pressure. We followed the global assembly of the galaxy, its
star-formation history and internal kinematics, and characterized
the masses, sizes, densities, dynamics and UV contribution of its
stellar clumps. The main results are:
\begingroup
\renewcommand{\labelitemi}{\textcolor{red}{\rule{1.2ex}{1.2ex}}}
\begin{itemize}

    \item The initially extended gas rapidly collapses into a
    compact, rotationally supported disk. Within approximately
    $20\,{\rm Myr}$, both gas and stars exhibit ordered rotation,
    with $V_{\rm rot}/\sigma_z>1$, while the stellar half-mass
    radius subsequently grows to $0.2-0.3\,{\rm kpc}$.

    \item The galaxy reaches a peak SFR of approximately
    $10$--$15\,\Msun\,{\rm yr}^{-1}$ and a final stellar mass of $4-7\times 10^8\ M_\odot$. Kinetic feedback reduces
    the cumulative stellar mass formed by $100\,{\rm Myr}$ by
    approximately $30$--$40\%$, but none of the explored feedback
    prescriptions prevents the initial disk fragmentation.

    \item Regions with $Q_{\rm gas}<1$ appear before the onset of
    star formation, and the first stellar structures form
    preferentially within these unstable regions. This temporal
    and spatial correspondence supports a Toomre-like
    gravitational fragmentation pathway.

    \item The resulting clumps have stellar masses of approximately
    $10^6$--$2\times10^8\,\Msun$, effective radii of
    $7-50\,{\rm pc}$, and mean stellar surface densities
    of $2\times10^3$--$3\times10^4\,
    \Msun\,{\rm pc}^{-2}$.

    \item Thermal supernova feedback produces results close to the
    no-feedback run because of rapid radiative losses. Kinetic
    feedback couples more effectively to the gas, lowers the gas
    mass retained within stellar-clump masks by approximately one
    order of magnitude, and suppresses the high-mass end of the
    clump mass function. Photoionization and direct radiation
    pressure produce smaller additional changes.

    \item Feedback shifts clumps towards smaller masses and
    radii approximately along loci of constant $\Sigma_\star$. It therefore regulates their subsequent growth and
    coherent extent more strongly than their initial formation.

    \item Even without feedback, clumps undergo substantial
    dynamical evolution. Massive systems can migrate inward,
    interact and coalesce, while tides and differential rotation
    produce stripping, deformation and dispersal. Gas removal by
    feedback makes stellar clumps more susceptible to these
    gravitational processes.

    \item In the fiducial \texttt{KnR\_fb} run, the fraction of
    stellar mass associated with clumps declines from $0.41$ at
    $20\,{\rm Myr}$ to $0.13$ at $100\,{\rm Myr}$. The
    corresponding intrinsic UV fraction decreases from $0.40$ to
    $0.10$, indicating that the declining UV contribution
    primarily reflects the decreasing stellar mass contained in
    clump masks.

    \item The simulated clumps overlap a substantial fraction of
    the mass--radius--surface-density parameter space occupied by
    observed $z>6$ clumps. The most compact and densest systems,
    particularly the Cosmic Gems, remain below the effective
    resolution of the present analysis.

\end{itemize}
\endgroup

These results indicate that the in-situ fragmentation of a
compact, gas-rich disk is a viable pathway for producing many of
the dense stellar structures observed in the early Universe.
Several questions nevertheless remain open. Simulations spanning
a broader range of halo properties and including cosmological
accretion and mergers are needed to establish how general this
pathway is. Higher spatial resolution, particle-based clump merger
trees and more complete feedback models -- including Ly$\alpha$
radiation pressure -- will be required to determine the clump birth
spectrum, lifetimes and bound fractions. Finally, comparing the
predicted Toomre scale directly with the initial masses and
separations of newly formed clumps, and following their evolution
in a cosmological tidal field, will be essential for testing
whether some survive as present-day compact stellar systems.

\begin{acknowledgements} 
We would like to thank A. Parichha, H. Rathore, Y. Nakazato, J. Rosdahl and R. Teyssier for useful feedback on various numerical aspects, and A. Adamo, M. Brada{\v{c}}, E. Vanzella for stimulating discussions. AF acknowledges support from the ERC Advanced Grant INTERSTELLAR H2020/740120. This research was supported by the Munich Institute for Astro- and Particle Physics (MIAPbP) of the DFG cluster of excellence "Origin and Structure of the Universe". Partial support from the Carl Friedrich von Siemens-Forschungspreis der Alexander von Humboldt-Stiftung Research Award is kindly acknowledged. This research was supported in part by grant NSF PHY-2309135 to the Kavli Institute for Theoretical Physics (KITP). We gratefully acknowledge computational resources of the Center for High Performance Computing (CHPC) at SNS. We acknowledge the use of plotting softwares: PYNBODY \citep{2013ascl.soft05002P} and yt \citep{2011ApJS..192....9T}.
\end{acknowledgements}

\bibliographystyle{aa_url}
\bibliography{ref}



\appendix
\section{Radiation pressure implementation in RAMSES-RT}

We only consider UV photons, no IR.
\subsection{Momentum Transfer Equation}
The gas momentum update is computed as:
\begin{equation}
\Delta p_{\rm gas,d} = f_{\rm boost} \sum_{i=1}^{5} F_{i,d} \cdot \Delta t \cdot (\kappa_{\rm abs,i} + \kappa_{\rm sc,i}) \cdot \frac{E_{i,\rm erg}}{c_{\rm RT}}
\end{equation}

where $d \in \{x, y, z\}$ and $c_{\rm RT} = 0.01 c$ is the reduced speed of light. $\kappa_{\rm abs,i} = \kappa_{\rm abs,i}^{\rm ion} + \kappa_{\rm abs,i}^{\rm dust} \quad [\text{s}^{-1}]$ is the total absorption rate.

\subsection{Photoionization Absorption}
From neutral species ionization:
\begin{equation}
\kappa_{\rm abs,i}^{\rm ion} = \sum_{j \in \{\text{H}_2, \text{HI}, \text{HeI}, \text{HeII}\}} n_j \cdot \sigma_{c,i,j} \cdot c_{\rm RT}
\end{equation}

where $\sigma_{c,i,j} = \text{group\_csn}(i,j) \cdot c_{\rm RT} \quad [\text{cm}^3\text{ s}^{-1}]$ and $n_j$ is neutral species densities $[\text{cm}^{-3}]$. The neutral species number densities are: $n_{\text{H}_2} = n_H \cdot x_{\text{H}_2}; n_{\text{HI}} = n_H \cdot x_{\text{HI}}; n_{\text{HeI}} = n_{\text{He}} \cdot x_{\text{HeI}}; n_{\text{HeII}} = n_{\text{He}} \cdot x_{\text{HeII}}$ where $n_{\text{He}} = 0.25 \cdot n_H \cdot Y/X$.

\subsection{Dust Absorption (Temperature-Independent)}
\begin{equation}
\kappa_{\rm abs,i}^{\rm dust} = \text{kappaAbs}_i \cdot \rho \cdot Z_{\odot} \cdot f_{\rm dust} \cdot c_{\rm RT}
\end{equation}

where: $f_{\rm dust} = 1 
; \rho = \frac{n_H \cdot m_H}{X} \quad [\text{g cm}^{-3}]; \text{kappaAbs}_i$ for all 5 groups are mentioned in Tab.\ref{tab:energy_groups}. We do not assume scattering contribution to momentum transfer, hence $\text{kappaSc}_i = 0 \text{ cm}^2\text{ g}^{-1} \quad \forall i \in [1,5]$

\section{Robustness and convergence of clump selection criteria}\label{app:select}
\begin{figure*}
    \centering
    \includegraphics[width=1\linewidth]{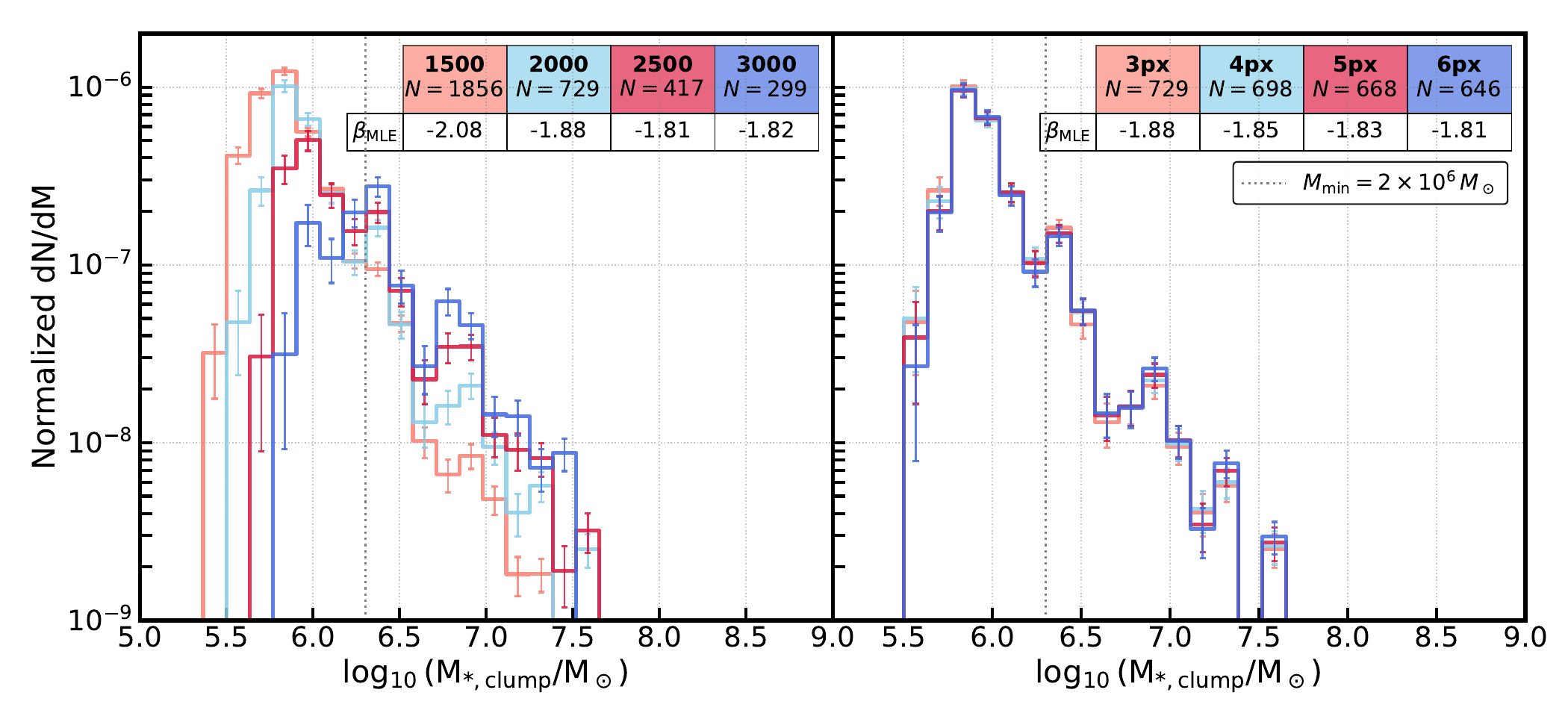}
    \caption{\textit{Left}: Stellar clump mass function for different stellar surface density thresholds of 1500 $\Msun \rm pc^{-2}$ (salmon), 2000 $\Msun \rm pc^{-2}$ (skyblue), 2500 $\Msun \rm pc^{-2}$ (crimson) and 3000 $\Msun \rm pc^{-2}$ (royalblue) for \texttt{KnR\_fb}. \textit{Right}: Stellar clump mass function for different choices of minimum peak-to-peak separation: 3 pixels (salmon), 4 pixels (skyblue), 5 pixels (crimson) and 6 pixels (royalblue) for \texttt{KnR\_fb}, where each pixel is $\approx$ 3.91 pc. The error bars are the Poisson uncertainty of the counts per bin. The total number of clumps (N), the $\beta$ values for unbinned MLE ($\beta_{{\rm MLE}}$) corresponding to each run are mentioned in the top left portion of the panel.}
    \label{fig:sigmathreshold}
\end{figure*}

To check the dependence of clump properties on the selection criteria, we compared the slope of CMF for four different stellar surface density threshold ($\Sigma_{\star,th}$: 1500, 2000, 2500 and 3000 $\Msun \rm pc^{-2}$) and four different choices of peak separation (11,9,7 and 5 cells) for the \texttt{KnR\_fb} run (see Fig. \ref{fig:sigmathreshold}). As expected, increasing $\Sigma_{\star,th}$ and the peak separation requirement reduced the number of detected clumps. Higher $\Sigma_{\star,th}$ selectively suppresses low mass clumps, while lower thresholds overpredict the number of smaller clumps. Despite this trend, all inferred CMF slopes using the MLE ($\beta_{\rm MLE}$) agree within $\approx$ 10\% within each other and with the observed slope $\beta_{\rm obs}= -1.89$ \citep{Claeyssens_Adamo_Kokorev_Furtak_Richard_Beauchesne_Dessauges-Zavadsky_Atek_Chisholm_Endsley_et_al_2026}. The dependence of the slope on peak separation is even weaker. The resulting slopes differ by only $\approx$ 2\% from each other and by $\approx$ 4\% from $\beta_{\rm obs}$. This tight convergence of $\beta_{\rm MLE}$ across a wide range of parameter choices validates the robust physical feature of the clump population rather than an artifact of parameter choices.

We also tested the numerical convergence by calculating the mass fraction of stars present in clumps over time in three different resolutions of 3.6 pc, 7.3 pc and 14.6 pc for \texttt{Th\_fb} (similar to \texttt{No\_fb} i.e. no feedback) using the same clump selection criterion as in Sec. \ref{sec:selectioncriteria}. In all three cases, the mass fraction reaches 0.65 by 40 Myr and then gradually goes to 0.60 by 100 Myr. 

\end{document}